\documentclass{pasj00}

\SetRunningHead{H.~ Imai et al.}{Water Masers in W51A}

\title{3-D Kinematics of Water Masers in the W51A Region}
\author{Hiroshi \textsc{Imai}\altaffilmark{1,2}
\thanks{present address, Joint Institute for VLBI in Europe, 
Postbus 2, 7990 AA Dwingeloo, the Netherlands}
Teruhiko \textsc{Watanabe}\altaffilmark{3}, 
Toshihiro \textsc{Omodaka}\altaffilmark{3}, 
Masanori \textsc{Nishio}\altaffilmark{3}, \\
Osamu \textsc{Kameya}\altaffilmark{1,2}, 
Takeshi \textsc{Miyaji}\altaffilmark{2, 4}, and 
Junichi \textsc{Nakajima}\altaffilmark{5}}
\altaffiltext{1}{Mizusawa Astrogeodynamics Observatory, 
National Astronomical Observatory, \\ Mizusawa, Iwate 023-0861}
\altaffiltext{2}{VERA Project Office, National Astronomical Observatory, 
Mitaka, Tokyo 181--8588}
\altaffiltext{3}{Department of Physics, Faculty of Science, 
Kagoshima University, Kagoshima 890-0065}
\altaffiltext{4}{Nobeyama Radio Observatory, 
National Astronomical Observatory, \\
Minamimaki, Minamisaku, Nagono 384-1305}
\altaffiltext{5}{Kashima Space Research Center, 
Communications Research Laboratory, \\ 
Kashima, Ibaraki 314-0012}
\email{(HI) imai@jive.nl}
\Received{2002/05/7}
\Accepted{2002/07/23}
\Published{2002/10/25}

\def\kms{~km~s$^{-1}$}
\def\h2o{H$_2$O}
\def\vlsr{$V_{\mbox{\scriptsize LSR}}$\ }
\def\etal{ et~al.\ }

\begin{document}

\maketitle

\begin{abstract}
We report proper motion measurements of water masers in the massive-star 
forming region W~51A and the analyses of the 3-D kinematics of the masers 
in three maser clusters of W51A (W51 North, Main, and South). In W~51 
North, we found a clear expanding flow that has an expansion velocity of 
$\sim$70\kms\  and indicates deceleration. The originating point of the 
flow coincides within 0\arcsec.1 with a silicon-monoxide maser source 
near the H{\rm II} region W~51d. In W51 Main, no systematic motion was 
found in the whole velocity range 
(158\kms\   $\leq\; $\vlsr  $<$ $-$58\kms) although a stream motion 
was reported previously in a limited range of the Doppler velocity 
(54\kms\   $\leq\; $\vlsr  $<$ 68\kms). Multiple driving sources of 
outflows are thought to explain the kinematics of W51 Main. In 
W51 South, an expansion motion like a bipolar flow was marginally visible. 
Analyses based on diagonalization of the variance-covariance matrix of 
maser velocity vectors demonstrate that the maser kinematics in W~51 
North and Main are significantly tri-axially asymmetric. We estimated a 
distance to W51 North to be 6.1$\pm$1.3 kpc on the basis of the 
model fitting method adopting a radially expanding flow. 
\end{abstract} 

\section{Introduction}
\label{sec:introduction}

Water maser emission is one of the most important phenomena in the study 
of star formation, often based on data obtained using very long baseline 
interferometry (VLBI) with high angular and velocity resolution (e.g., 
\cite{rei81,eli92}). Analyses of spatial positions, Doppler velocities, and 
proper motions of individual maser features with a typical size of 1 AU \citep
{rei81} have revealed the 3-D gas kinematics around young stellar objects 
(YSOs) (e.g., \cite{gen81a}a, hereafter G81; \cite{gen81b}b; \cite
{sch81}, hereafter S81; \cite{rei88,gwi92,cla96,fur00,ima00}, hereafter 
Paper {\rm I}; \cite{tor01}). In practice, details of the gas kinematics are 
complicated but seem to depend mainly on the evolutionary status of YSOs. 
Water maser sources sometimes enable measurement of the internal motions 
of giant molecular clouds by measuring relative bulk motions between 
clusters of water masers that are very close to each other (Paper {\rm I}; 
\cite{tor01}). Such bulk motions may be owe to, e.g.,  propagation of 
shock layers from newly-formed H{\rm II} regions, cloud contraction by 
the self-gravitation due to the huge mass of a giant molecular cloud. 

The massive-star forming region W~51A contains at least five independent 
clusters of water masers: W51 North, West, Main, and two clusters in South 
(e4 and e3). All of the maser clusters are independently associated with 
H{\rm II} regions and dense and young cloud cores exhibiting several 
species of molecular emission (\cite{gen78, dow79}; G81; S81; \cite{gau87, 
gau93, zha95}, 1997; \cite{zha98,lep98}, hereafter LLD; \cite{eis02}, 
hereafter EGHMM). In W51N, silicon-monoxide masers have been 
detected (\cite{mor92}; EGHMM). Proper motions of these water masers 
were measured two decades ago (G81; S81). However, details of the 
kinematics are still obscured because of limited numbers of measured proper 
motions ($\leq$30, S81; G81). In W51 North, a separation motion has been 
recognized between two dominant "sub-clusters" of maser features (see 
Sect.\ \ref{sec:results-summary}), so-called the "NW Cluster" and the 
"Dominant Center Reference Cluster" (S81). EGHMM reconfirmed 
the similar separation motion in the patterns of the sub-clusters, which 
are similar to bow shocks, by comparing the patterns observed in 1998 
with those of S81. In W51 Main, LLD found a stream from the 
"Double-Knot" sub-cluster in the south-east direction. The observed 
velocity range, however, was very limited 
(54\kms\   $<$ \vlsr  $<$ 68\kms). Throughout the entire range 
($-$50\kms\   $\leq$ \vlsr  $\leq$ 130\kms) the kinematics are 
predominated by random motions (G81). In other maser clusters, their 
kinematics have never been well understood. 

Here, we report monitoring observations of the W~51A water masers with 
the Japanese domestic VLBI network (J-Net) \footnote
{J-Net includes the 45-m telescope of Nobeyama Radio Observatory (NRO), 
which is a branch of the National Astronomical Observatory, an interuniversity 
research institute operated by the Ministry of Education, Culture, Sports, 
Science and Technology.}
\citep{omo94}. Sect.\ 
2 describes the VLBI observation and data reduction. Sect.\ 3 summarizes 
the revealed 3-D kinematics of the individual clusters of water masers.
Sect.\ 4 discusses the origins and related issues of the maser kinematics. The 
estimation of the distance to W~51A and relative 3-D bulk motions among the 
maser clusters are also described.

\section{Observations and Data Reduction}
The VLBI observations were made at five epochs for a period of 8 months 
in 1999, using three or four J-Net telescopes. Table \ref{tab:status} gives the 
status of the J-Net observations. The spatial resolution of our observations 
was typically $~$3 milliarcseconds (mas). W51 North/West, Main, and two 
regions of W51 South, close to the H{\rm II} regions W51 e4 or e8 and e3 
(\cite{gau93,zha98}), are located within a 70\arcsec-field. We observed 
simultaneously the four clusters of masers within a single antenna beam 
(the minimum size of 72\arcsec\ [FWHM] at 22 GHz in the array telescopes). 
At each epoch, an observation was made for 10-11 hours including scans 
of calibrators (NRAO~530 and JVAS~2145$+$067. The received data were 
recorded with the VSOP terminal \citep{kaw94} in a single base band channel 
with a band width of 16 MHz, which corresponds 
to a velocity coverage of 216\kms\ (158\kms\   $\leq\; $\vlsr  $<$ $-$58\kms). 

The data correlation was made with the Mitaka FX correlator \citep{chi91}. 
Correlations were performed four times by shifting the phase-tracking centers 
to the locations of the above four clusters of masers. An average correlation time 
was set to 0.5 s for W51 North and 1 s for W51 Main and two regions of W~51 
South. The correlation outputs consisted of 1024 velocity channels with a 
velocity spacing of 0.21\kms\ each. 

Data reduction was made with the NRAO AIPS package using normal 
procedures (e.g., \cite{dia95}). Fringe fitting and self-calibration procedures 
were performed for a Doppler velocity channel including a bright maser spot 
(velocity component) in W51 North, also given in Table \ref{tab:status}. 
The solutions were applied to all of the data, then maser maps were made 
for all of the maser clusters. The typical size of the synthesized beam was 
3 mas in the five observations (see Table \ref{tab:status}). The relative 
position accuracy of a maser spot ranged over 0.02--0.6 mas depending on a 
signal-to-noise ratio and spatial structure of the spot. Identification of a 
water maser feature was made in the same procedure shown in several 
previous papers (e.g., Paper {\rm I}). A relative position accuracy of a maser 
feature was ranged over 0.05--0.6 mas. Relative proper motions were measured 
for maser features identified in at least two of the five epochs. 

\section{Results}


\subsection{Summary of proper motion measurements}
\label{sec:results-summary}

\begin{figure}
  \begin{center}
    \FigureFile(85mm, 170mm){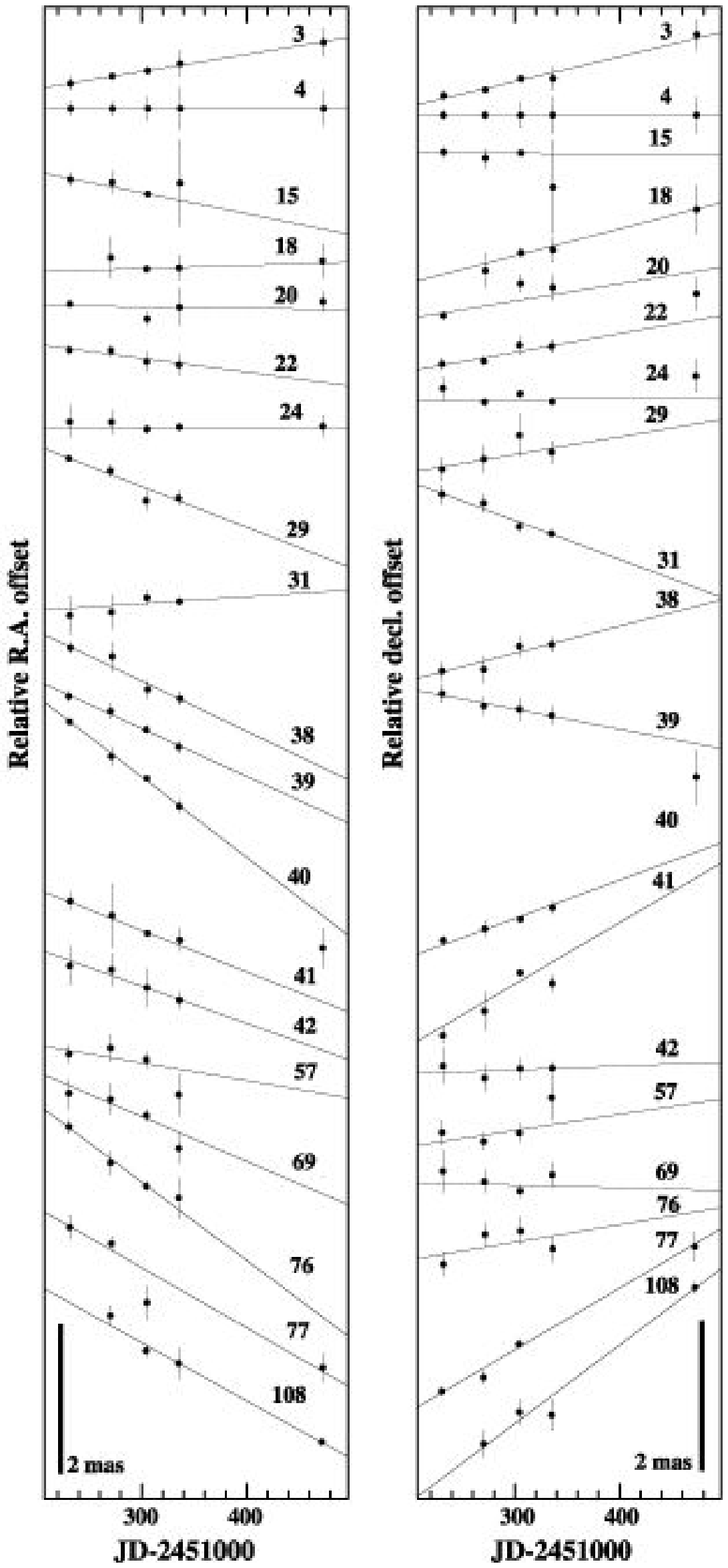}
  \end{center}
\caption{Observed relative proper motions of water maser features in 
W51 North. Only the maser features detected at four or all five 
epochs are presented here. A number added for each proper motion 
shows the assigned one after the designated name form 
``W51N:I2002". A solid line indicates a least-square-fitted line 
assuming a constant velocity motion.}
\label{fig:PM-W51N}
\end{figure}

Figure \ref{fig:PM-W51N} shows several examples of measured relative 
proper motions of water maser features. Maser features fundamentally 
seem to move with constant velocities. The deviations from fit lines 
assuming constant velocity motions are within several tenths of a 
milliarcsecond. Some features have a large deviation from the fit lines 
because they are located together with other features within a small 
range, 1 mas in space and 1\kms\   in velocity, in which we were not 
able to correctly trace the same feature from one epoch to another. 

Tables \ref{tab:pmotionsN}, \ref{tab:pmotionsM}, and \ref{tab:pmotionsS} 
give parameters of maser features with measured proper motions. The 
numbers of measured proper motions were 123, 48, and 10 in W51 North, 
Main, and South (e4), respectively, which are larger than those of previous 
observations for W51 North by S81 and for W51 Main and South (e4) by 
G81. An important difference between the previous and the present 
measurements is the difference in time separations between the 
successive observing epochs: two years and, at minimum, only one month, 
respectively. 

Even with a much shorter time separation of our observations, it has not 
been possible to measure proper motions in a large fraction of detected 
maser features ($>$ 50 \%) mainly because of growth and decay of maser 
features among the observing epochs. These results imply that many 
of the maser features have lifetimes shorter than 1--2 months. 
Unfortunately, no maser proper motion has been identified in the W 51 
South (e3) region.

Usually, each of water maser clusters consists of several groups of maser 
features with a size of 100--1000 AU. In this paper, we define such a 
group of features as a "sub-cluster".


\subsection{W51 North}
\label{sec:W51N}
\subsubsection{Overview}
\label{sec:W51Noverview}

\begin{figure*}
  \begin{center}
    \FigureFile(160mm, 160mm){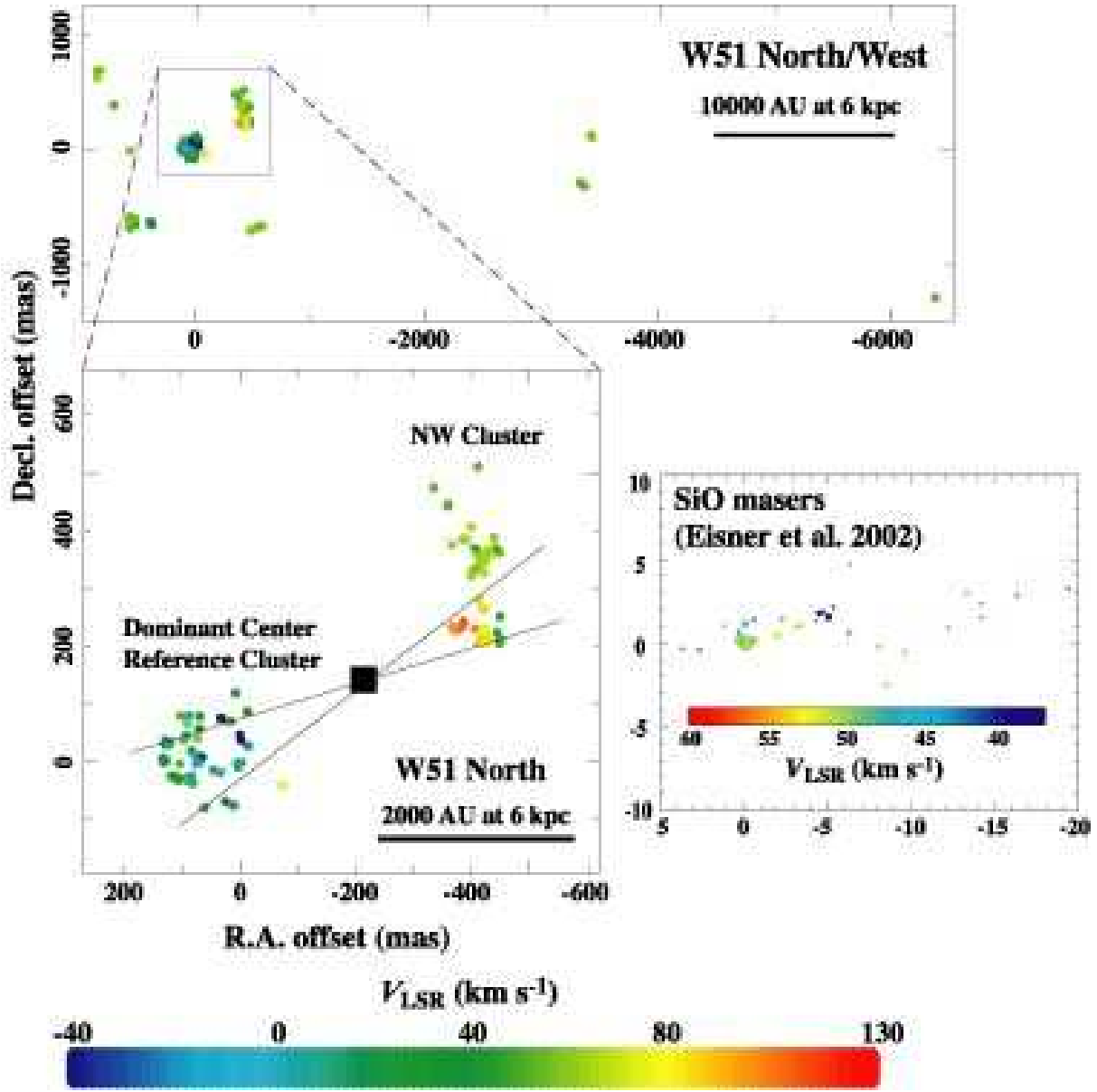}
  \end{center}
  \caption{Doppler velocity distribution of water masers in the W51 North 
and West regions. A filled rectangle in the lower panel indicates the location 
of Silicon monoxide maser emission in W51N observed by EGHMM, which 
was estimated in the procedure mentioned in the main text. The 
distribution of the SiO masers obtained by EGHMM is also shown in the 
sub-frame. The opening angle of the SiO maser distribution has been 
extrapolated to the scale of the \h2o maser emission (solid lines), which is 
the same as that of EGHMM.} 
\label{fig:W51N-color}
\end{figure*}

Figure \ref{fig:W51N-color} presents the angular distribution of water 
masers in W51 North and West, which revives those found by previous 
observations (\cite{gen78,dow79}; S81; EGHMM). Figure \ref
{fig:3D-W51N} presents the detailed 3-D motions of water masers around 
the two dominant sub-clusters, the red-shifted (north--west) and the 
blue-shifted (south--east, Dominant Center Reference) sub-clusters, 
which clearly exhibit a bipolar expanding flow. We found a "bow shock" 
pattern in the SE sub-cluster, which opens toward the NW sub-cluster 
and has been found since 1977 (S81; EGHMM). On the basis of the 
locations of the two sub-clusters and this bow-shock pattern, we 
estimated a location of SiO maser emission in W51N on our map with 
an uncertainty of less than 100 mas, whose position was measured by 
EGHMM with respect to the water masers with an uncertainty of less 
than 50 mas. The originating point of the outflow seems to be located 
around the middle of the two maser sub-clusters and to be roughly 
coincident with the location of the SiO maser. While only a separation 
motion between the two sub-clusters has been confirmed by S81 and 
EGHMM, the present result reveals that the individual sub-clusters 
themselves are also expanding. 

On the other hand, motions of maser features far from the two 
sub-clusters do not exhibit any systematic motion but random motions 
with velocities up to 100\kms. Some peculiar motions are found in the 
two dominant sub-clusters and are not due to misidentification of the 
proper motions. Thus, although the expanding flow has been visible, 
the kinematics of the W51 North region is heavily disordered dynamically 
by fast random motions. The details of the maser kinematics around 
W51 West were also obscured because only two proper motions were 
measured. It is difficult to compare the maser distribution with that of 
S81 because of too sparse time separation ($\sim$20 yr). 

\subsubsection{Model fitting the maser kinematics}
\label{sec:model-fit}
In order to estimate kinematical parameters of the expanding flow and 
a distance to W51N, we made model-fitting analyses for the 3-D motions 
of maser features. The procedure was performed on the basis of the 
least-square fitting of the observed kinematics to a radially-expanding 
flow model and in almost the same way as that applied to the W3 IRS~5 
water masers (Paper {\rm I}), which is not repeated here. One difference 
is only the assumed speed of the radial expansion of a maser feature {\it i}, 
$V(i)$, as a function of distance of a maser feature from the originating 
point of the outflow, $r_{i}$, which is expressed more simply as 
$V_{\mbox{exp}}(i)\;=\;V_{0}(r_{i}/r_{0})^{\alpha}$, and where $V_{0}$ is 
an expansion velocity at a unit distance $r_{0}$, $\alpha$ is a power-law 
index indicating the apparent acceleration of the flow. We made the 
fitting step-by-step, excluding maser features having unreliably 
large positive or negative expansion velocities or distances from the outflow 
origin. After such several iterations, we used 68 proper motion data and 
obtained best solutions that are given in Table \ref{tab:w51model-fit}. 

Figure \ref{fig:3D-W51N} shows an estimated position of the outflow 
origin in the maser motions (plus), which coincides with the location of 
the SiO maser emission (filled square, EGHMM) within the position uncertainly 
($\sim$100 mas). A systemic line-of-sight velocity of the flow is almost equal 
to that of the ambient molecular cloud ($\simeq$56\kms, e.g., S81; \cite
{cox87, rud90, zha95, zha98,oku01}) and roughly coincident with that of 
the SiO maser emission ($\simeq$47\kms) within a velocity width of the 
cloud ($\simeq$26\kms). 

\begin{figure*}
  \begin{center}
    \FigureFile(170mm, 170mm){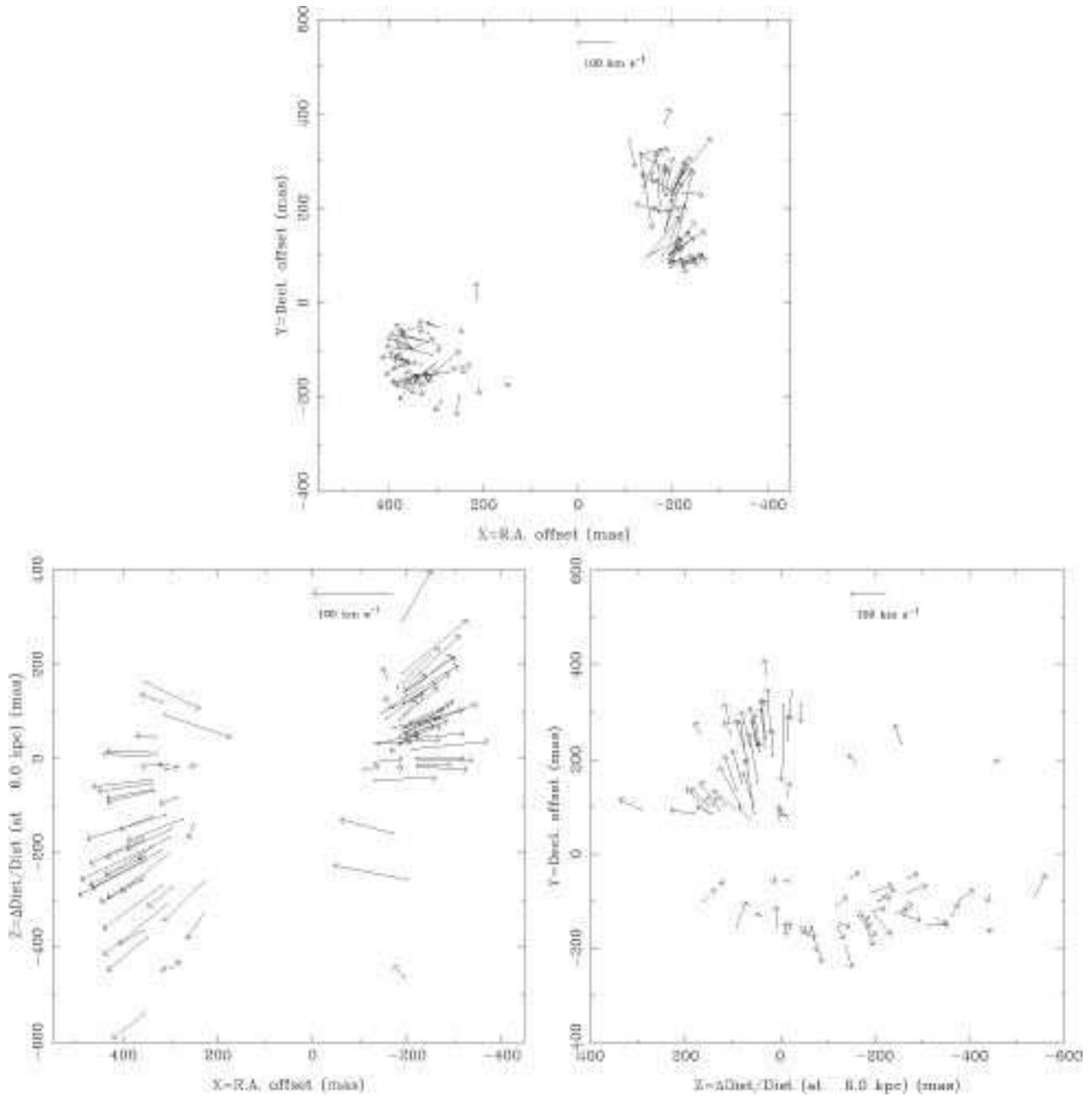}
  \end{center}
\caption{Estimated three-dimensional positions and motions of water maser 
features in W51 North. The positions and motions are with respect to those 
of the outflow origin determined by a model fit. The position of an arrow 
indicates that of a maser feature. Direction and length of an arrow indicates 
the direction and the magnitude of the maser motion, respectively. {\it Top}: 
Front view (\mbox{$XY$}-plane) of the positions and motions. {\it Bottom left}: 
Top view (\mbox{$XZ$}-plane). {\it Bottom right}: East-side view 
(\mbox{$ZY$}-plane).}
\label{fig:W51N-3DV}
\end{figure*}

\begin{figure}
  \begin{center}
    \FigureFile(85mm, 60mm){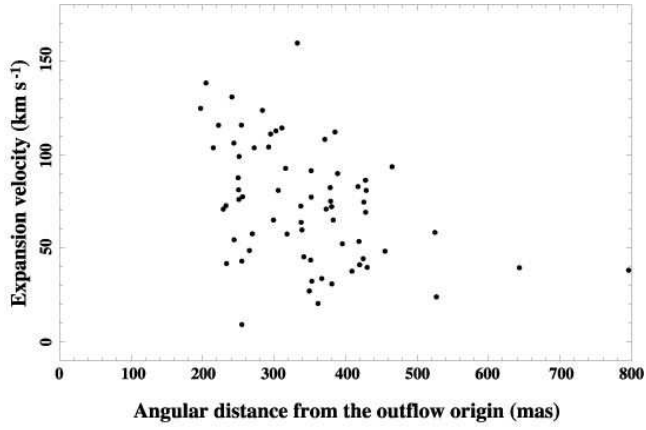}
  \end{center}
\caption{Radial expansion velocity of a water maser feature against 
a radial distance from the outflow origin in W51 North after a model fit. 
According to the best-fit model, the expansion velocity is 
$V_{\mbox{exp}}$\mbox{$\simeq$94\kms}\  at \mbox{$r=$ 0\arcsec.25} 
and $V_{\mbox{exp}}$\mbox{$\simeq$55\kms}\  at \mbox{$r=$ 0\arcsec.45}. }
\label{fig:expansion}
\end{figure}

Figure \ref{fig:W51N-3DV} presents the maser feature motions projected 
onto three different planes. The maser kinematics indicate no rotation of 
the expanding flow, suggesting that ballistic motions predominate the 
kinematics. The best fit model and an expansion velocity plot against 
distance from the outflow origin (Figure \ref{fig:expansion}) indicate that 
the expanding flow decelerates in the water maser region ($r=$ 200--500 
mas or 1200--3000 AU from the outflow origin), where the expansion 
velocity decreases from $\simeq$90\kms\   to $\simeq$50\kms. This is a 
controversial case against these expanding flows that apparently exhibit 
the accelerations in water maser kinematics (Orion KL, \cite{gen81b}b; 
W~49N, \cite{gwi92}; W3 IRS~5, Paper {\rm I}). 

\subsection{W51 Main}
\label{sec:W51M}



\begin{figure*}
  \begin{center}
    \FigureFile(170mm, 140mm){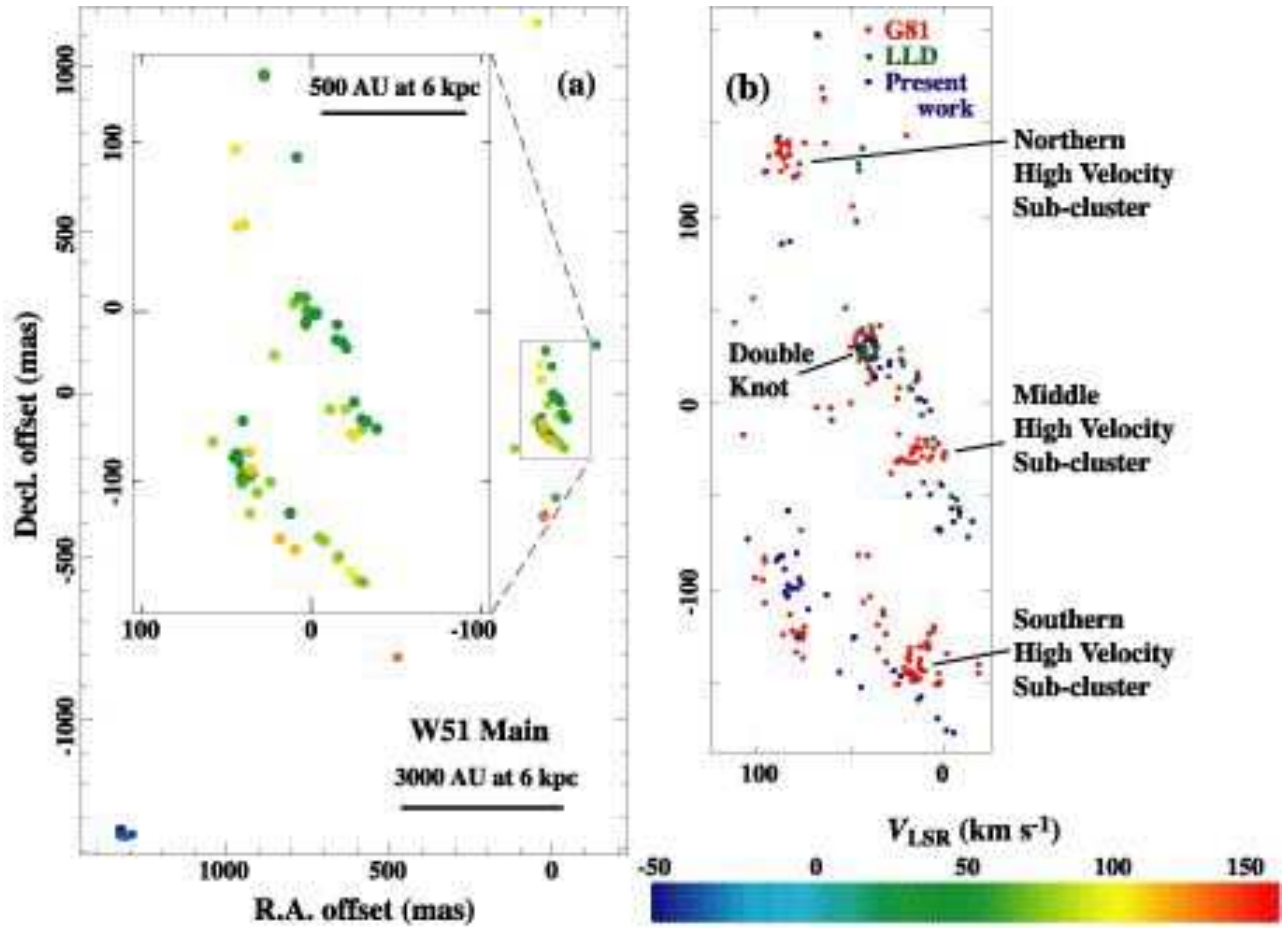}
  \end{center}
  \caption{Same as Fig.~2 but in the W51 Main region. (a) Water masers 
found in the present work. (b) Maser distribution obtained by superposition 
of maps of G81 (observed in 1977--1979, red filled circle), LLD (observed in 
1994, greed filled circle) and the present work (observed in 1999, blue filled 
circle). The coordinate of G81 was adopted. The double knot at (45, 25) in 
unit of mas was used as reference of the superposition.}
\label{fig:W51M-color}
\end{figure*}

Figure \ref{fig:W51M-color}a presents the angular distribution of water 
masers in W51 Main, which also revives those found by previous 
observations (\cite{gen78, gen79}; G81) shown in Figure \ref
{fig:W51M-color}b. Four maser sub-clusters have been identified by G81: 
"Double Knot", "Middle High Velocity Cluster", "Northern High Velocity 
Cluster", and "Southern High Velocity Cluster", all of which seem to be 
stable for at least 20 years. 

LLD identified a "cocoon" in the Double Knot, which is more clearly 
seen by superposing three maps of G81, LLD, and the present work 
around the coordinate (45, 25) in unit of mas in Figure \ref
{fig:W51M-color}b. Assuming a distance to W51M of 6 kpc, this cocoon 
has an inner and an outer radii of approximately 12 AU and 60 AU, 
respectively. A rotation of the cocoon has been also identified by LLD, 
but it was not found in the present work, probably because of too small 
a number of maser features detected around the cocoon. 

On the other hand, the remaing maser sub-clusters have large position 
offsets up to 20 mas (120 AU at a distance of 6 kpc) between the previous 
and the present maps. Likely these are not true motions of sub-clusters 
but the "Christmas tree" effect due to appearance and disappearance of 
maser features during 20 years. Unlike the bow shock pattern seen in 
W~51N, no clear feature alignment was found except for the cocoon 
mentioned above. 

\begin{figure}
  \begin{center}
    \FigureFile(85mm, 140mm){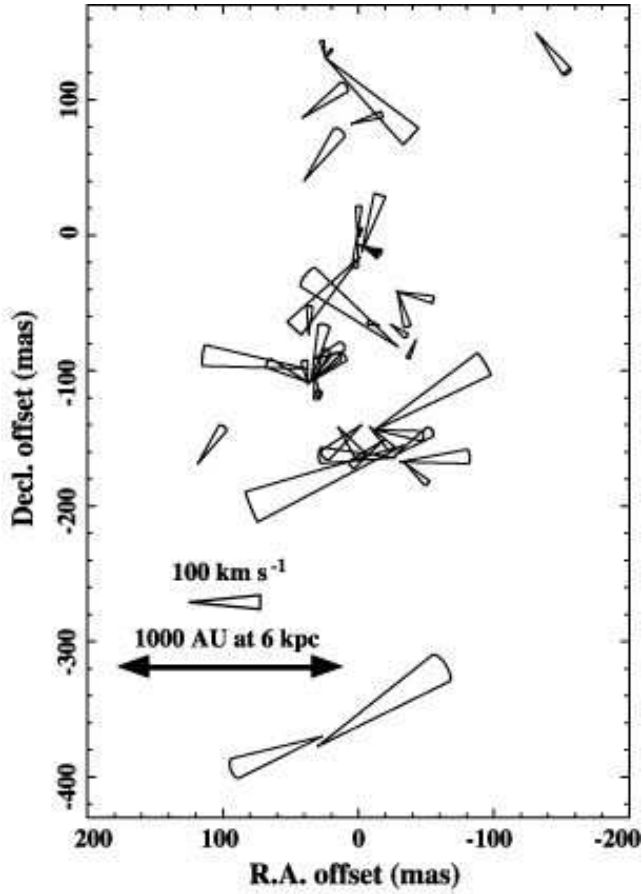}
  \end{center}
  \caption{Same as Fig.~3 but for W51 Main. An inclination of a maser 
motion is calculated with respect to the systemic Doppler velocity of the parent 
molecular cloud (\mbox{$\simeq$55\kms}).} 
\label{fig:3D-W51M}
\end{figure}

Figure \ref{fig:3D-W51M} presents the detailed 3-D motions of water 
masers around the four maser sub-clusters. Most of all maser features 
with measured proper motions are red-shifted with respect to the 
systemic velocity of this region (50--60\kms, e.g., G81; \cite
{cox87,rud90, zha95, ho96, zha98, oku01}). 
Looking at the whole Doppler-velocity range (158\kms\  $\leq\; $\vlsr $<$ 
$-$58\kms), the kinematics of water masers is apparently random. 
On the other hand, LLD measured 26 proper motions of water 
masers with a one-month time baseline and found a stream in the 
SW direction from the cocoon in a limited range of the Doppler velocity 
(54\kms\  $\leq\; $\vlsr $<$ 68\kms). We note that the Southern High 
Velocity Cluster seems to independently have an expanding flow. Maser 
features on the north--east side of this sub-cluster (around the coordinate 
[40, $-$100] in unit of mas in Figure \ref{fig:3D-W51M}) have proper 
motions toward the cocoon with velocities of 40--80\kms, probably which 
are not a flow contracting toward the cocoon but an expanding flow from 
the point around the coordinate (20, $-$140) in unit of mas in Figure \ref
{fig:3D-W51M}. The systemic Doppler velocity of the candidate originating 
this expanding flow is expected to be around \vlsr $\sim$90\kms\ on the 
basis of the mean Doppler velocity of maser features in the sub-cluster. 

\subsection{W51 South (e4)}
\label{sec:W51S}

\begin{figure}
  \begin{center}
    \FigureFile(85mm, 60mm){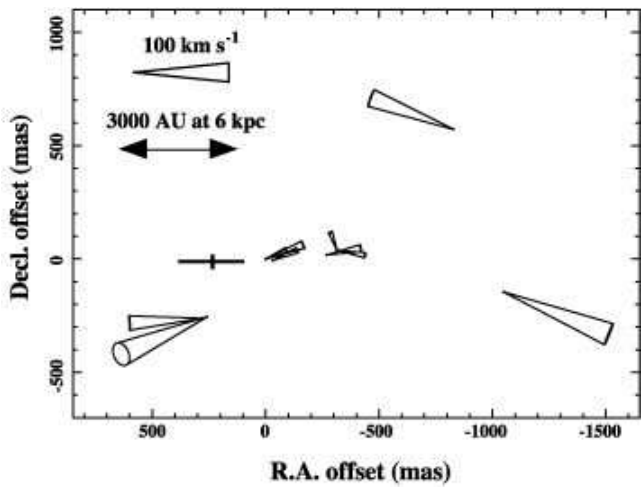}
  \end{center}
  \caption{Same as Fig.~3 but for W51 South. A plus at 
(230 mas, \mbox{$-$}10 mas) indicates an estimated position of the expanding 
flow in W51 South.} 
\label{fig:3D-W51S}
\end{figure}

Figure \ref{fig:3D-W51S} presents the 3-D motions of water maser 
features in W51 South, close to the 3.6-cm continuum source W51 e4 
\citep{gau93} and 2-mm continuum source W~51 e8 \citep{zha98}. The 
distribution of water masers in this region seems to have been roughly 
stable and aligned in the north--west to south--east direction (G81). We 
found that the water masers exhibit marginally a Doppler-velocity gradient 
along the maser alignment and a systematic separation motion between 
the red-shifted and the blue-shifted masers with bipolarity. We attempted 
a model fitting for the water masers using only a model such as that applied 
as Step 1 for W51 North water masers, in which we estimated only 
the originating point of the outflow (see Sect.\ \ref{sec:model-fit} and 
Paper {\rm I}). Table \ref{tab:w51smodel-fit} gives its solution. 

\section{Discussion}


\subsection{Distance to the W~51A region}
\label{sec:distance}

The distance to the W51A region (G~49.5$-$0.4) has been adopted to be 
$\sim$7 kpc on the basis of the statistical parallax of water-maser proper 
motions measured by G81 (c.f.\ S81). Because of an increase in measured 
proper motions and of reliable fitting of the maser kinematics to an 
expanding flow model in the present paper, a distance of $\sim$6 kpc 
should be adopted. 

\begin{figure*}
  \begin{center}
    \FigureFile(140mm, 140mm){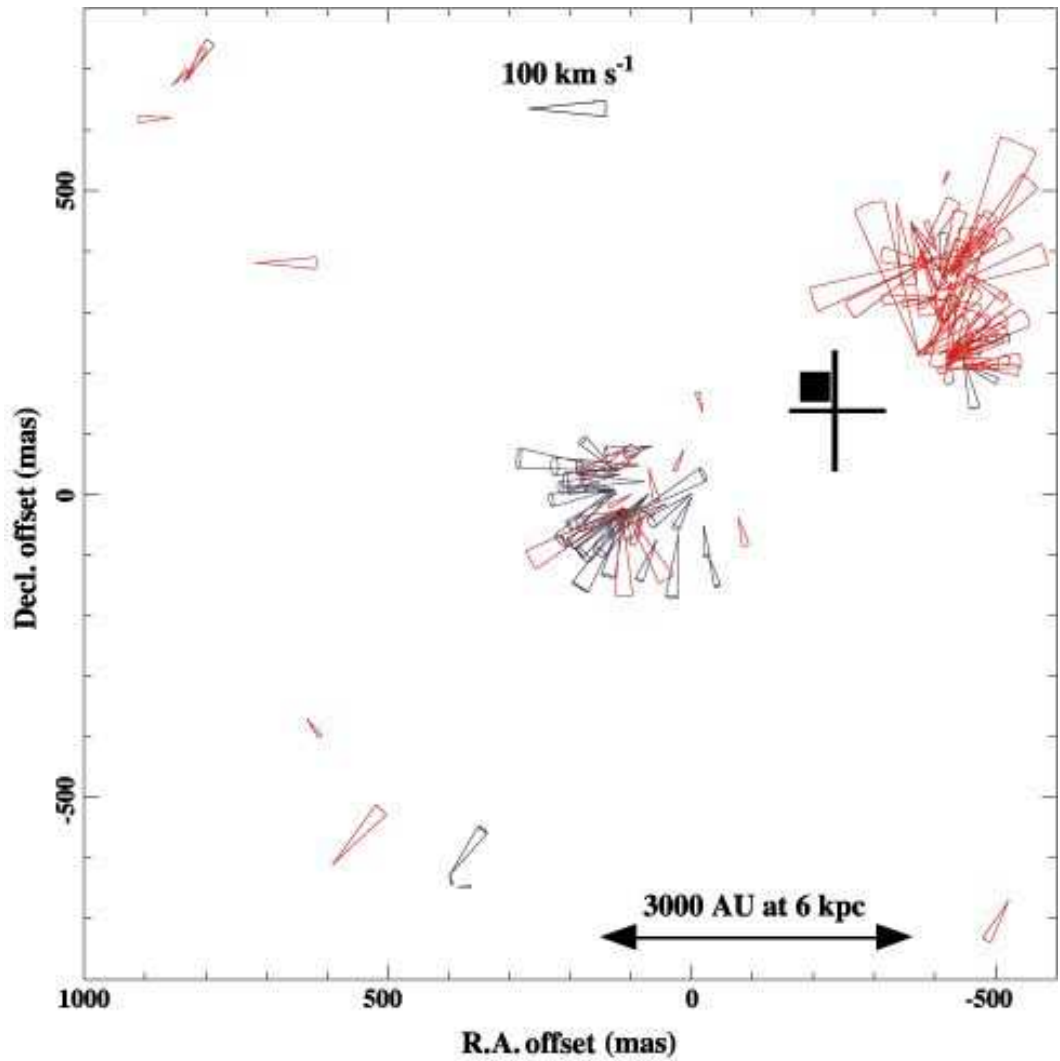}
  \end{center}
  \caption{3-D motions of water maser features in W51 North. 
A position, opening direction, and length of a cone indicate the position 
of a maser feature and the direction and amplitude of its proper 
motion, respectively. An inclination of a maser motion is calculated 
with respect to the systemic Doppler velocity of the parent molecular cloud 
(\mbox{$\simeq$56\kms}). Red and blue cones indicate receding and 
approaching features, respectively. An origin of the map is located at 
the position-reference maser feature. A plus at (\mbox{$-$}230 mas, 130 mas) 
indicates an estimated position of the expanding flow in W51 North. 
A filled rectangle indicates the same as that in Figure 2.} 
\label{fig:3D-W51N}
\end{figure*}

As described in Sect.~\ref{sec:model-fit} and Table \ref{tab:w51model-fit}, 
we obtained a distance value of 6.1$\pm$1.3 kpc from a model fitting for 
the kinematics of W51N. The uncertainty permits a distance value of 
7 kpc. When adopting 7 kpc, however, the model fitting puts the 
originating source of the W51N outflow at a point with a NW offset of 
$~\sim$0\arcsec.1 from that when adopting 6 kpc. As seen in Figures 
\ref{fig:W51N-color} and \ref{fig:3D-W51N}, this offset is inconsistent 
with the suggestion that the originating point should coincide to the 
location of SiO masers (EGHMM). Moreover, although both of the NW and 
SE sub-clusters exhibit deceleration, unlike a simple deceleration as shown 
in Figure \ref{fig:expansion}, they are not aligned on a simple one. 

We also attempted the statistical parallax method and obtained a distance 
value of 7.2$\pm$1.7 kpc but when including all measured proper motions 
in the W~51 North and West regions. When using only proper motions in 
the two dominant sub-clusters of W51N, an unreliably small value 
($\sim$4 kpc) was obtained. We obtained a distance value of 8.5$\pm$3.3 
kpc from the kinematics of W51M. As described in Sect.\ \ref{sec:vvcm} and 
Figures \ref{fig:3D-W51N} and \ref{fig:W51N-3DV}, the 3-D maser 
kinematics are heavily biased. Therefore, it is difficult to obtain a reliable 
distance value using the statistical parallax. Here we adopt the distance 
obtained by the model fitting. 

On the other hand, the distance to G49.5$-$0.4 has been estimated to be 
5.5 kpc on the basis of a 'far' kinematical distance adopting the Galactic 
constants: $R$\solar $=$ 8.5 kpc and $\Theta$\solar $=$ 220\kms\ (c.f., 
\cite{cra78, dow80}), which is quite consistent with that we have obtained 
from the model fitting. Note that, for some of the Galactic water maser 
sources, the distances obtained from their water maser kinematics 
coincide quite well with their kinematical distances (e.g., W3~IRS~5, 
Paper {\rm I}). 

\subsection{Objective analyses of the maser kinematics}
\label{sec:vvcm} 

In order to model-independently deal with the kinematics of water masers 
in W~51 North and Main, we performed diagonalization for the velocity 
variance--covariance matrices obtained from velocity vectors of maser 
features (c.f., \cite{blo00}). Table \ref{tab:vvcm} gives the obtained 
eigenvalues (velocity dispersions) and their corresponding eigenvectors 
after the diagonalization. For example, for a bipolar outflow, an eigenvector 
corresponding to the largest eigenvalue indicates a major axis of the flow 
and others have two almost equal eigenvalues much smaller than the largest 
one. In the cases of both W51 North and Main, the obtained three 
eigenvalues differ from each other by factors larger than two; the maser 
kinematics are not isotropic but tri-axially asymmetric. Although the magnetic 
drag of bipolar flows might be applicable to explain such asymmetry \citep
{blo00}, geometrical effects should be taken into account. 

A major eigenvector with the largest corresponding eigenvalue of the 
W51N kinematics (an inclination of 16\arcdeg$\pm$11\arcdeg and 
a position angle of $-$41\arcdeg$\pm$68\arcdeg) is almost parallel to the 
major axis of the W51N outflow, which can be confirmed in Figures \ref
{fig:3D-W51N} and \ref{fig:W51N-3DV}. EGHMM also obtained similar 
angles of the outflow axis on the basis of their observations of 
the W51N SiO masers. Having two eigenvalues much larger than the one 
for the W51N kinematics can be explained by the outflow having a large 
opening angle, which is much larger than that estimated from SiO masers 
($\sim$25\arcdeg, EGHMM) and by the flow being strongly blocked at 
some points. 

A geometrical effect such as that expected in W51N is also expected in 
the W51M kinematics. This can explain why most water masers are 
red-shifted with respect to the systemic velocity (\vlsr$=$ 50--60\kms, 
see Sect.\ \ref{sec:W51M}) when supposing that the driving source of 
the water masers is located behind a dense molecular cloud core that 
blocks the flow material approaching us. This hypothesis is supported by 
the fact that the eigenvector with the smallest eigenvalue is almost along 
the line-of-sight. Note that possible multiple outflows along the sky plane 
as mentioned in Sect.\ \ref{sec:W51M} can also create such tri-axially 
asymmetric kinematics. 

\subsection{Origins of the maser kinematics}
\label{sec:origin}

Water masers in massive-star forming regions are associated with outflows 
from newly-formed massive stars. Therefore, we can estimate the locations 
of young massive stars as originating points of the outflows and, in some 
cases, elucidate the evolutionary statuses of the young stars and kinematical 
structures of the outflows themselves as implied in Sect.\ \ref{sec:introduction}. 

For the W51N flow, EGHMM proposed that the SiO maser should be 
associated with a conical flow that rotates (counter clockwise) about the 
axis of the expansion motion. On the other hand, we did not find such a 
helical motion in the \h2o maser kinematics as mentioned in Sect.\ \ref
{sec:model-fit}. Although \h2o and SiO maser flows are driven from a 
common young star, the associating flows are likely streaming 
independently. Such a hybrid system was also proposed for the 
Orion-KL outflows \citep{gre98} and, therefore, might be a common 
characteristic of outflows from massive young stars. On the other hand, 
the W51N maser kinematics are heavily biased due to possible blocking of 
the flow as mentioned in Sect.\ \ref{sec:vvcm}. Adjacent young massive stars 
and their embedding dense cloud cores are expected as candidates of such 
blocking objects because the W51N region is composed of several massive 
young stars within a very small volume ($<$ 0.1 pc) (e.g., \cite{wat98, oka01}). 
They may also create observed fast ($\leq$100\kms) random motions in the 
maser kinematics in fields far from the flow origin and between the cloud 
cores. It is difficult to create such random motions by means of shocks in 
the interstellar medium, which are triggered by adjacent H{\rm II} regions, 
because the H{\rm II} regions themselves will expand much more slowly 
($V_{\mbox{exp}}\leq$ 30\kms). 

The existence of several massive stars in the W~51M is expected as 
mentioned in Sect.\ \ref{sec:W51M}. In fact, although a duration of 
\h2o maser activity is quite short ($<$10$^{5}$ yr, \cite{gen77}), multiple 
driving sources in single water maser clusters have been found in some 
massive-star forming regions (e.g.\ W~3 IRS~5, Paper {\rm I}; Cep~A, 
\cite{tor01}). These imply that a mass infall rate in a star-forming region 
exhibiting such multiple outflows should be much higher than 
10$^{-3}$ $M$\solar yr$^{-1}$. This lower limit of the rate is calculated 
using an assumed total mass of the star cluster larger than 100 $M$\solar (in 
the case of W51N, \cite{oka01}) divided by the time scale within which 
this mass infalls before \h2o masers are quenched by photoionization 
of \h2o molecules due to the oldest massive star in the star cluster. 

\subsection{Relative 3-D motions of water maser clusters in W~51A}
\label{sec:rmotions} 

In principle, it is possible to measure relative and systemic bulk motions 
between clusters of water masers in the same manner as the "in-beam 
astrometry" because they were observed simultaneously in a single 
beam of each telescope. In the present paper, the coordinates of all 
maser maps were precisely fixed with respect to the position-reference 
maser feature in W51N (W51N: I2002 {\it 4}). Table \ref{tab:rmotion} 
gives measured relative bulk motions. 

In practice, it is difficult to precisely estimate such bulk motions of the 
maser clusters because these motions are heavily disordered by 
random motions. Even for W51N, whose maser kinematics are relatively 
well known, the uncertainty of its systemic bulk motion is quite large 
($\sim$40\kms, see Table \ref{tab:w51model-fit}) compared with the 
typical velocity width of a molecular cloud ($<$ 10\kms). For W51M, 
only the mean proper motion of the W51M kinematics is observable and 
adopted as a systemic bulk motion of W51M; of course its uncertainty 
is also quite huge. 

Adopting the systemic bulk motions of W51 North and South obtained 
from model fittings (Table \ref{tab:w51model-fit} and \ref
{tab:w51smodel-fit}) and the mean maser motion of W51 Main as the 
systemic bulk motion, we found a bulk motion separating W51 North 
and Main/South by $\sim$130\kms\ (c.f.\ \cite{ima01}). At first, 
the separation motion driven by pressures by the adjacent HII regions is 
considered. In fact, there are bright HII regions that are located 0.9 pc 
west from W51 Main/South (W~51 IRS~1) and 0.1 pc north--east from 
W51 North (W~51 IRS~1/d) (e.g.\ \cite{gen82, gol94}). On the other 
hand, the mechanical luminosity required for the bulk motions is 
calculated assuming a spherically expanding flow sweeping molecular 
clouds (maser regions) that have a density of molecular hydrogen, 
$n_{\mbox{\scriptsize H$_{2}$}}\sim$ 10$^{6}$ cm$^{-3}$ (e.g., 
\cite{eli92}), in an expansion velocity to create the bulk motion, 
$v_{\mbox{exp}}\simeq$ 130/2\kms, at a distance of $r\simeq$ 0.9 pc. 
However, the obtained luminosity, $>$ 10$^{7}$ L\solar, is too huge to 
be produced only by stellar clusters in the W~51A region. The biases in 
the obtained systemic bulk velocities of the clusters are apparently most 
likely to be generated as discussed in Sect.\ \ref{sec:vvcm}. Some artificial 
effects on the coordinate drifts between the maser clusters are also 
expected: for example, unknown large offsets of telescope positions 
due to lack of geodetic VLBI observations with J-Net will result in 
the failure to resolve 2$\pi$-n ambiguity and to connect fringe-phases 
through each of the observations. 

\section{Summary}

We have obtained data on the 3-D kinematics of three clusters of water 
masers in W~51A with VLBI monitoring observations composed of five 
epochs with an 8-month time baseline. The main conclusions of this 
paper are as follows. 

1. In W51N, a bipolar outflow is clearly exhibited by the maser kinematics, 
modelled using a radially expanding flow model. The position of the driving 
source of the outflow coincides with that of the SiO maser source within 
0\arcsec.1. Expansion velocity is consistent with that obtained by EGHMM 
on the basis of the change in maser distribution for 15 years but decreasing 
with distance from the driving source from 90\kms\ to 50\kms. 

2. In W51M, random motions predominate the maser kinematics. Multiple 
outflows are considered. Unlike W51N, no systematic change in the 
maser distribution has been seen during 20 years due to the "Christmas 
tree effect" for maser appearance and disappearance. 

3. In W51S, a bipolar outflow was found marginally. The driving source of 
the outflow has a large offset ($>$ 1\arcsec) from both of the continuum 
sources W51 e4 and e8. 

4. The maser kinematics in W~51A are heavily biased due to possible 
concentration of young massive stars and their embedding clouds 
that destroy symmetry or systematic motion of the outflows. Duration 
of active massive-star formation in each of the maser clusters is shorter 
than 10$^{5}$ yr in W51A, which is suggested by the fact that the \h2o 
masers associated with the different driving sources of outflows are 
simultaneously observed. 

5. The distance to W51A was estimated to be 6.1$\pm$1.3 kpc on the basis 
of the model fitting method applied to the maser kinematics of W51N. On 
the estimated distance, the kinematical model for the W51N outflow is 
consistent with that previously proposed on the basis of VLA observations 
(EGHMM) and exhibits simple deceleration in the flow. This distance value is 
smaller than that previously adopted ($\sim$7 kpc) but consistent with 
the kinematical distance ($\sim$5.5 kpc). 

6. Measurements of relative 3-D bulk motions between the maser clusters 
have been attempted and showed an apparent separation motion 
between W~51 North and Main/South. To elucidate the true motions, 
random motions and biases in the maser kinematics should be carefully taken 
into account. Reliable kinematical models are indispensable to 
find accurately systemic bulk motions of maser clusters. 

\bigskip
We gratefully acknowledge all staff members and students who have 
helped in array operation and in data correlation of the J-Net. We also 
thank Drs. T.~Sasao, Y.~Asaki, and M.~Miyoshi for providing the previous 
J-Net data. We are also grateful to Drs. M.~J.~Reid, L.~J.~Greenhill, 
T.~Liljestr\"om, and S.~Inutsuka for valuable comments and 
Dr. J.~A.~Eisner for providing information on SiO masers in W51 North. 
H. I. was financially supported by the Research Fellowship of the Japan 
Society of the Promotion of Science for Young Scientist.

\clearpage

\clearpage

\clearpage

\clearpage

\clearpage

\clearpage

\clearpage 
\setlength{\baselineskip}{6ex}

\begin{table*}[ht]
\caption{Status of the telescopes, data reduction, and resulting performances in 
the individual epochs of the J-Net observations.}\label{tab:status}
\begin{center}
\scriptsize

\begin{tabular}{lccccccccc} \hline \hline
& Epoch in & & & Reference & 1-$\sigma$ level & Synthesized & & & \\
Observation & the year & Duration & Used    
& velocity\footnotemark[2] & noise 
& beam\footnotemark[3] 
& \multicolumn{3}{c}{Features\footnotemark[4]} \\
code & 1999 & (hr) & telescopes\footnotemark[1]  
& (\kms) & (Jy beam$^{-1}$) & (mas)
& N & M & S \\ \hline 
j99052 \dotfill & February 21 & 11 & M, b, k & 37.74 & 9.3 
& 7.4$\times$2.4, $-$38.4$^{\circ}$ & 137 & 51 & 22 \\
j99092 \dotfill & April 2 & 10 & M, b, n\footnotemark[5], k & 34.67 & 3.8 
& 8.8$\times$2.9, $-$51.6$^{\circ}$ & 100 & 50 & 17 \\
j99126 \dotfill & May 6 & 10 & M, b, n, k & 35.72 & 2.9 
& 8.1$\times$3.0, $-$41.3$^{\circ}$ & 171 & 88 & 23 \\
j99157 \dotfill & June 6 & 10 & M\footnotemark[6], b\footnotemark[6]$^{,}$
\footnotemark[7], n\footnotemark[6], k\footnotemark[6] & 35.51 & 4.0 
& 7.2$\times$2.6, $-$54.2$^{\circ}$ & 61 & 33 &12 \\
j99294 \dotfill & October 21 & 11 & M, b, k & 53.95 & 22.0 
& 7.6$\times$2.4, $-$38.2$^{\circ}$ & 47 & 25 & 6 \\ \hline
\end{tabular}

\end{center}
\footnotemark[1] Attending telescopes; M: the 10-m telescope at Mizusawa, 
b: the 34-m telescope at Kashima, n: the 45-m telescope at Nobeyama, 
k: the 6-m telescope at Kagoshima. \\
\footnotemark[2] Velocity channel used for the phase reference in data 
reduction. \\
\footnotemark[3] The synthesized beam made in natural weight; 
major and minor axis lengths and position angle. \\
\footnotemark[4] Number of detected maser features in the W51A region 
(North, Main, and South [e4]). \\
\footnotemark[5] Ceasing operation for 7 hr due to strong winds. \\
\footnotemark[6] High system temperature due to bad weather conditions. \\
\footnotemark[7] Data heavily affected by unstable local-frequency signals. 
\end{table*}

\begin{table*}[p]
\caption{Parameters of the water maser features identified by 
proper motion toward W51 North.} \label{tab:pmotionsN}
{\scriptsize
\begin{tabular}{l@{ }r@{ \ }rr@{ \ }rr@{ \ }rrrr@{ \ }r@{ \ }r@{ \ }r@{ \ }r} \hline \hline
 & \multicolumn{2}{c}{Offset\footnotemark[2]}
 & \multicolumn{4}{c}{Proper motion\footnotemark[3]}
 & \multicolumn{2}{c}{Radial motion\footnotemark[4]}
 & \multicolumn{5}{c}{Peak intensity\footnotemark[5]} \\                                         
Maser Feature\footnotemark[1] & \multicolumn{2}{c}{(mas)} 
 & \multicolumn{4}{c}{(mas yr$^{-1}$)}
 & \multicolumn{2}{c}{(km s$^{-1}$)}
 & \multicolumn{5}{c}{(Jy beam$^{-1}$)} \\
 & \multicolumn{2}{c}{\ \hrulefill \ } 
 & \multicolumn{4}{c}{\ \hrulefill \ } 
 & \multicolumn{2}{c}{\ \hrulefill \ } 
 & \multicolumn{5}{c}{\ \hrulefill \ } \\                                        
(W51N:I2002) & R.A. & decl. 
 & $\mu_{x}$ & $\sigma \mu_{x}$ & $\mu_{y}$ 
 & $\sigma \mu_{y}$ & $V_{z}$ & $\Delta V_{z}$
 & j99052 & j990092 & j99126 & j99156 & j99294 \\  \hline
  1   \ \dotfill \  &     79.47 &   $-$22.41 &    1.00 &   0.72 &    0.59 &   0.76
 & $-$60.28 &   2.85
  &    ...     &       19.97 &       46.68 &    ...     &    ...       \\     
  2   \ \dotfill \  &     88.94 &     47.52 &    1.92 &   0.73 &    1.29 &   0.51
 & $-$54.49 &   1.33
  &    ...     &       12.82 &       37.50 &        7.03 &    ...       \\     
  3   \ \dotfill \  &     89.08 &     47.20 &    0.84 &   0.29 &    1.24 &   0.28
 & $-$52.22 &   2.66
  &      892.23 &      130.49 &      211.22 &        8.83 &      156.40   \\     
  4   \ \dotfill \  &      0.00 &      0.00 &    0.00 &   0.36 &    0.00 &   0.35
 & $-$47.71 &   1.85
  &      110.23 &        8.71 &        9.17 &        2.05 &      144.14   \\     
  5   \ \dotfill \  &     72.09 &    $-$0.90 &    1.09 &   1.10 &  $-$1.47 &   1.32
 & $-$46.45 &   1.89
  &    ...     &        5.99 &       12.98 &    ...     &    ...       \\     
  6   \ \dotfill \  &     72.39 &    $-$1.23 &    0.38 &   1.82 &  $-$1.38 &   2.37
 & $-$44.81 &   2.74
  &      100.20 &        5.41 &    ...     &    ...     &    ...       \\     
  7   \ \dotfill \  &    119.36 &     31.27 &    0.00 &   1.31 &    1.52 &   1.11
 & $-$39.86 &   1.69
  &    ...     &        8.23 &       10.32 &    ...     &    ...       \\     
  8   \ \dotfill \  &    129.20 &      7.92 &    1.00 &   0.46 &    0.80 &   0.92
 & $-$34.65 &   2.25
  &       45.38 &        5.57 &       30.25 &    ...     &    ...       \\     
  9   \ \dotfill \  &    125.34 &     31.14 &    0.68 &   2.38 &    0.76 &   2.45
 & $-$34.24 &   1.79
  &    ...     &        3.56 &        5.00 &    ...     &    ...       \\     
 10   \ \dotfill \  &    124.67 &     38.42 &    0.59 &   1.31 &    1.26 &   1.33
 & $-$32.96 &   1.89
  &      190.68 &        5.24 &    ...     &    ...     &    ...       \\     
 11   \ \dotfill \  &    125.62 &     38.57 &  $-$0.70 &   0.98 &    2.14 &   0.66
 & $-$31.83 &   1.54
  &    ...     &        2.36 &    ...     &        5.58 &       30.60   \\     
 12   \ \dotfill \  &    125.45 &     30.24 &  $-$4.20 &   1.10 &    1.65 &   1.60
 & $-$31.41 &   0.74
  &    ...     &    ...     &        3.03 &    ...     &       14.90   \\     
 13   \ \dotfill \  &    129.33 &      8.30 &    0.55 &   2.05 &  $-$0.98 &   2.42
 & $-$31.31 &   1.47
  &    ...     &        2.73 &       13.37 &        2.83 &    ...       \\     
 14   \ \dotfill \  &    125.69 &     38.30 &  $-$0.38 &   1.65 &    2.72 &   1.27
 & $-$30.37 &   1.54
  &    ...     &        1.71 &       17.69 &        3.83 &    ...       \\     
 15   \ \dotfill \  &      0.63 &    $-$7.08 &  $-$1.03 &   0.50 &  $-$0.02 &   0.48
 & $-$28.94 &   1.37
  &       89.45 &        6.60 &       20.87 &        2.48 &    ...       \\     
 16   \ \dotfill \  &     79.72 &     19.64 &    1.73 &   1.41 &    1.40 &   1.50
 & $-$28.66 &   1.06
  &    ...     &        9.01 &        3.69 &    ...     &    ...       \\     
 17   \ \dotfill \  &     82.41 &   $-$25.44 &  $-$4.68 &   2.69 &    3.00 &   2.10
 & $-$26.16 &   0.84
  &    ...     &        1.98 &        4.03 &    ...     &    ...       \\     
 18   \ \dotfill \  &    130.33 &      6.28 &    0.07 &   0.50 &    1.36 &   0.67
 & $-$25.81 &   1.80
  &    ...     &        2.77 &       12.12 &        3.95 &       25.10   \\     
 19   \ \dotfill \  &     43.26 &   $-$17.39 &    0.71 &   1.56 &    0.15 &   1.33
 & $-$25.33 &   0.94
  &       44.58 &        6.17 &    ...     &    ...     &    ...       \\     
 20   \ \dotfill \  &    126.16 &      4.00 &  $-$0.12 &   0.21 &    0.88 &   0.29
 & $-$24.58 &   1.10
  &       61.40 &    ...     &       15.29 &        2.48 &       46.81   \\     
 21   \ \dotfill \  &     66.70 &     78.80 &    0.24 &   1.81 &    1.06 &   0.99
 & $-$24.21 &   1.37
  &      247.27 &       10.30 &    ...     &    ...     &    ...       \\     
 22   \ \dotfill \  &     66.88 &     78.89 &  $-$0.69 &   0.47 &    0.92 &   0.38
 & $-$23.14 &   1.89
  &      151.73 &       29.84 &       66.22 &        9.21 &    ...       \\     
 23   \ \dotfill \  &    109.27 &   $-$25.40 &  $-$2.19 &   2.14 &  $-$1.49 &   2.44
 & $-$20.22 &   1.37
  &      188.58 &        2.48 &    ...     &    ...     &    ...       \\     
 24   \ \dotfill \  &    125.98 &      0.70 &  $-$0.09 &   0.29 &    0.08 &   0.27
 & $-$20.14 &   2.15
  &       64.66 &       16.81 &       91.50 &       14.81 &       46.37   \\     
 25   \ \dotfill \  &   $-$19.76 &   $-$53.79 &  $-$1.91 &   1.20 &  $-$0.05 &   1.05
 & $-$20.10 &   0.94
  &      155.72 &        7.09 &    ...     &    ...     &    ...       \\     
 26   \ \dotfill \  &    109.32 &   $-$25.55 &  $-$1.13 &   0.25 &  $-$1.30 &   0.60
 & $-$16.36 &   2.32
  &      649.10 &       29.12 &       76.13 &    ...     &    ...       \\     
 27   \ \dotfill \  &     21.99 &   $-$66.78 &  $-$1.54 &   0.71 &  $-$1.56 &   0.78
 & $-$15.71 &   1.26
  &       96.70 &       13.81 &       37.69 &    ...     &    ...       \\     
 28   \ \dotfill \  &    128.27 &    $-$0.23 &  $-$0.70 &   0.26 &    1.64 &   0.35
 & $-$14.45 &   2.11
  &    ...     &    ...     &      139.10 &    ...     &       37.50   \\     
 29   \ \dotfill \  &    398.09 &  $-$628.58 &  $-$1.94 &   0.34 &    0.92 &   0.73
 & $-$14.43 &   2.85
  &      457.59 &       40.26 &       94.65 &       15.65 &    ...       \\     
 30   \ \dotfill \  &    109.33 &   $-$25.85 &  $-$0.36 &   0.57 &  $-$0.65 &   0.47
 & $-$14.34 &   2.10
  &      427.39 &       45.80 &        7.99 &    ...     &    ...       \\     
 31   \ \dotfill \  &    104.82 &   $-$27.55 &    0.33 &   0.66 &  $-$1.78 &   0.34
 & $-$12.80 &   3.74
  &      339.00 &       37.72 &      168.41 &       30.10 &    ...       \\     
 32   \ \dotfill \  &    397.93 &  $-$628.08 &  $-$3.37 &   2.23 &    3.41 &   2.12
 & $-$12.46 &   1.37
  &    ...     &        4.83 &        9.90 &    ...     &    ...       \\     
 33   \ \dotfill \  &   $-$28.11 &  $-$103.92 &  $-$2.30 &   5.16 &  $-$0.05 &   0.82
 & $-$11.59 &   1.68
  &    ...     &       24.90 &    ...     &        7.20 &    ...       \\     
 34   \ \dotfill \  &    111.24 &   $-$24.56 &    0.21 &   1.28 &  $-$0.61 &   1.08
 &  $-$9.19 &   1.26
  &       40.45 &    ...     &       10.52 &    ...     &    ...       \\     
 35   \ \dotfill \  &    110.37 &   $-$25.53 &    0.62 &   1.22 &  $-$0.33 &   0.65
 &  $-$8.04 &   2.00
  &    ...     &       27.12 &    ...     &       13.27 &    ...       \\     
 36   \ \dotfill \  &    109.97 &   $-$25.66 &    0.85 &   0.94 &  $-$0.07 &   0.83
 &  $-$6.85 &   3.26
  &    ...     &    ...     &      240.26 &       27.67 &    ...       \\     
 37   \ \dotfill \  &     58.31 &   $-$77.80 &  $-$1.05 &   0.99 &  $-$0.53 &   0.47
 &  $-$5.76 &   1.33
  &      193.45 &    ...     &       47.93 &        5.65 &    ...       \\     
38   \ \dotfill \  &    384.15 &  $-$648.39 &  $-$2.39 &   0.40 &    1.38 &   0.51
 &  $-$4.68 &   1.15
  &      190.35 &       21.65 &       74.08 &       11.67 &    ...       \\     
 39   \ \dotfill \  &  $-$446.62 &    224.33 &  $-$2.37 &   0.36 &  $-$1.02 &   0.55
 &  $-$3.99 &   4.11
  &     1100.97 &      304.45 &      899.17 &       90.83 &    ...       \\     
 40   \ \dotfill \  &  $-$451.21 &    251.39 &  $-$3.98 &   0.18 &    1.90 &   0.31
 &  $-$2.81 &   1.26
  &     1437.13 &      100.98 &      233.55 &       43.96 &      133.00   \\     
 41   \ \dotfill \  &  $-$409.00 &    368.84 &  $-$2.04 &   0.55 &    3.05 &   0.48
 &  $-$1.34 &   1.37
  &       52.66 &        8.49 &       38.54 &       11.88 &    ...       \\     
 42   \ \dotfill \  &     86.73 &   $-$31.02 &  $-$1.82 &   0.86 &    0.22 &   0.75
 &  $-$0.37 &   1.26
  &       64.60 &       25.37 &       82.28 &        9.91 &    ...       \\     
 43   \ \dotfill \  &  $-$451.77 &    210.57 &  $-$3.26 &   0.38 &    0.61 &   0.53
 &  $-$0.02 &   1.96
  &     3118.90 &       65.87 &       27.59 &    ...     &    ...       \\     
 44   \ \dotfill \  &    110.14 &   $-$26.31 &  $-$3.80 &   3.89 &  $-$1.77 &   3.19
 &    0.51 &   1.47
  &    ...     &    ...     &       40.85 &        2.40 &    ...       \\     
 45   \ \dotfill \  &     78.56 &   $-$35.88 &  $-$0.65 &   0.61 &    0.16 &   0.61
 &    0.77 &   1.75
  &      152.09 &    ...     &       21.67 &        3.70 &    ...       \\     
 46   \ \dotfill \  &     14.00 &     70.74 &  $-$1.43 &   0.98 &    0.43 &   0.96
 &    1.45 &   1.05
  &       43.04 &    ...     &       18.58 &    ...     &    ...       \\     
 47   \ \dotfill \  &  $-$450.39 &    213.09 &  $-$4.25 &   1.77 &    1.04 &   1.92
 &    1.66 &   1.15
  &    ...     &       11.76 &        6.40 &    ...     &    ...       \\     
 48   \ \dotfill \  &  $-$449.59 &    204.77 &  $-$3.88 &   1.98 &    1.83 &   2.18
 &    2.18 &   1.26
  &    ...     &        4.83 &       15.28 &    ...     &    ...       \\     
 49   \ \dotfill \  &     92.38 &   $-$29.12 &  $-$1.85 &   0.57 &    1.10 &   0.54
 &    2.35 &   1.19
  &       63.23 &       10.37 &       14.29 &    ...     &    ...       \\     
 50   \ \dotfill \  &    715.06 &    380.37 &  $-$4.54 &   5.44 &    1.20 &   2.53
 &    2.41 &   1.15
  &    ...     &    ...     &       32.03 &        6.85 &    ...       \\     
 51   \ \dotfill \  &    115.64 &   $-$23.59 &    0.28 &   0.85 &    2.48 &   0.91
 &    2.59 &   1.26
  &       35.71 &    ...     &       23.06 &    ...     &    ...       \\     
 52   \ \dotfill \  &     82.42 &   $-$28.60 &  $-$2.36 &   1.86 &    1.13 &   1.25
 &    3.42 &   0.77
  &       21.90 &        2.81 &        4.71 &    ...     &    ...       \\     
 53   \ \dotfill \  &    632.26 &  $-$372.77 &  $-$2.43 &   1.56 &    0.51 &   2.10
 &    3.56 &   0.84
  &    ...     &       16.75 &    ...     &        3.82 &    ...       \\     
 54   \ \dotfill \  &  $-$361.44 &    446.11 &  $-$4.45 &   1.65 &  $-$1.28 &   1.67
 &    4.23 &   1.68
  &    ...     &       35.57 &       72.44 &    ...     &    ...       \\     
 55   \ \dotfill \  &    100.03 &     80.76 &    0.41 &   1.71 &  $-$0.12 &   1.72
 &    4.35 &   1.33
  &       67.69 &       11.20 &        5.57 &    ...     &    ...       \\     
 56   \ \dotfill \  &    853.80 &    673.65 &  $-$2.50 &   0.88 &    1.88 &   0.86
 &    4.50 &   0.98
  &    ...     &       20.32 &       44.09 &        5.77 &    ...       \\     
 57   \ \dotfill \  &    100.48 &    $-$2.43 &  $-$0.86 &   0.71 &    0.82 &   0.88
 &    4.55 &   0.94
  &       38.64 &       17.82 &       27.56 &        2.19 &    ...       \\     
 58   \ \dotfill \  &  $-$413.52 &    510.54 &  $-$2.31 &   0.93 &    1.91 &   0.61
 &    5.13 &   1.05
  &    ...     &    ...     &       43.84 &    ...     &      170.90   \\     
 59   \ \dotfill \  &  $-$403.21 &    336.24 &  $-$1.60 &   0.31 &    3.17 &   0.27
 &    5.18 &   1.26
  &    ...     &    ...     &    ...     &       16.93 &      570.01   \\     
 60   \ \dotfill \  &  $-$360.93 &    444.40 &  $-$3.01 &   1.15 &  $-$7.00 &   0.86
 &    5.74 &   2.00
  &      812.00 &    ...     &       61.31 &    ...     &    ...       \\     
 61   \ \dotfill \  &  $-$361.75 &    445.23 &  $-$4.62 &   0.49 &    1.00 &   0.63
 &    6.30 &   1.41
  &    ...     &    ...     &       69.61 &       12.74 &       93.88   \\     
 62   \ \dotfill \  &  $-$361.34 &    447.69 &  $-$2.58 &   1.00 &  $-$2.90 &   5.46
 &    6.36 &   2.74
  &     7974.74 &       73.72 &    ...     &    ...     &    ...       \\     
 63   \ \dotfill \  &    589.86 &  $-$610.02 &  $-$3.99 &   5.29 &    3.76 &   5.17
 &    6.51 &   0.84
  &    ...     &        7.00 &       17.93 &    ...     &    ...       \\     
 64   \ \dotfill \  &  $-$337.35 &    474.86 &  $-$2.45 &   2.51 &  $-$2.29 &   2.39
 &    6.65 &   1.37
  &    ...     &    ...     &       64.81 &       13.47 &    ...       \\     
 65   \ \dotfill \  &  $-$450.98 &    362.90 &  $-$3.86 &   0.47 &    0.29 &   0.45
 &    7.10 &   1.05
  &    ...     &    ...     &       32.71 &    ...     &      115.27   \\     
 66   \ \dotfill \  &  $-$521.11 &  $-$671.25 &  $-$0.95 &   3.16 &  $-$0.46 &   2.85
 &    9.17 &   1.05
  &    ...     &    ...     &       44.25 &        6.58 &    ...       \\     
 67   \ \dotfill \  & $-$3400.86 &    124.02 &  $-$0.32 &   0.93 &  $-$0.29 &   1.63
 &    9.46 &   1.05
  &       37.96 &    ...     &        7.20 &    ...     &    ...       \\     
 68   \ \dotfill \  & $-$3402.38 &    124.99 &  $-$0.83 &   1.22 &    0.00 &   1.31
 &   11.25 &   1.26
  &       20.03 &    ...     &        3.68 &    ...     &    ...       \\     
 69   \ \dotfill \  &     71.00 &     40.89 &  $-$2.18 &   0.87 &  $-$0.13 &   0.93
 &   11.39 &   1.21
  &       29.73 &        2.44 &       11.16 &        1.98 &    ...       \\     
 70   \ \dotfill \  &     87.59 &     66.49 &  $-$0.82 &   0.79 &    0.63 &   0.84
 &   11.41 &   1.26
  &       34.61 &        1.85 &        6.58 &    ...     &    ...       \\     
 71   \ \dotfill \  &  $-$440.58 &    386.60 &  $-$5.12 &   2.94 &    4.88 &   2.35
 &   11.56 &   0.63
  &    ...     &        2.69 &        3.49 &    ...     &    ...       \\     
 72   \ \dotfill \  &    834.84 &    680.64 &  $-$3.84 &   0.72 &    3.05 &   0.73
 &   12.01 &   1.54
  &    ...     &       13.92 &       21.10 &        5.37 &    ...       \\     
 73   \ \dotfill \  &    114.77 &   $-$24.05 &    2.27 &   3.83 &  $-$1.13 &   4.89
 &   12.15 &   1.15
  &    ...     &    ...     &        6.75 &        2.27 &    ...       \\     
 74   \ \dotfill \  &  $-$413.23 &    355.80 &  $-$2.63 &   0.64 &    4.36 &   1.07
 &   13.06 &   1.19
  &       60.11 &        9.36 &        8.52 &    ...     &    ...       \\     
 75   \ \dotfill \  &    828.79 &    684.06 &  $-$2.52 &   0.55 &    2.69 &   0.39
 &   13.82 &   1.12
  &       75.30 &        6.20 &       12.94 &    ...     &    ...       \\     
\hline
\end{tabular}
}
\end{table*}

\addtocounter{table}{-1}
\begin{table*}[ht]
\caption{(Continued)}
{\scriptsize
\begin{tabular}{l@{ }r@{ \ }rr@{ \ }rr@{ \ }rrrr@{ \ }r@{ \ }r@{ \ }r@{ \ }r} \hline \hline
& \multicolumn{2}{c}{Offset\footnotemark[2]}
 & \multicolumn{4}{c}{Proper motion\footnotemark[3]}
 & \multicolumn{2}{c}{Radial motion\footnotemark[4]}
 & \multicolumn{5}{c}{Peak intensity\footnotemark[5]} \\
Maser Feature\footnotemark[1] & \multicolumn{2}{c}{(mas)} 
 & \multicolumn{4}{c}{(mas yr$^{-1}$)}
 & \multicolumn{2}{c}{(km s$^{-1}$)}
 & \multicolumn{5}{c}{(Jy beam$^{-1}$)} \\
 & \multicolumn{2}{c}{\ \hrulefill \ } 
 & \multicolumn{4}{c}{\ \hrulefill \ } 
 & \multicolumn{2}{c}{\ \hrulefill \ } 
 & \multicolumn{5}{c}{\ \hrulefill \ } \\                                        
(W51N:I2002) & R.A. & decl. 
 & $\mu_{x}$ & $\sigma \mu_{x}$ & $\mu_{y}$ 
 & $\sigma \mu_{y}$ & $V_{z}$ & $\Delta V_{z}$
 & j99052 & j990092 & j99126 & j99156 & j99294 \\  \hline
76   \ \dotfill \  &  $-$401.66 &    321.88 &  $-$3.83 &   0.56 &    0.85 &   0.76
 &   14.12 &   1.21
  &       37.37 &        7.06 &       10.14 &        2.81 &    ...       \\     
 77   \ \dotfill \  &    835.32 &    680.28 &  $-$3.19 &   0.35 &    3.17 &   0.29
 &   14.40 &   2.42
  &      140.49 &        6.36 &       12.34 &    ...     &       49.54   \\     
78   \ \dotfill \  &  $-$427.69 &    358.49 &  $-$2.99 &   0.92 &    3.05 &   1.06
 &   14.79 &   0.98
  &       18.33 &       11.57 &       11.98 &    ...     &    ...       \\     
 79   \ \dotfill \  &    857.86 &    619.66 &  $-$0.45 &   2.83 &    1.33 &   2.69
 &   14.90 &   1.58
  &    ...     &        2.22 &        7.51 &    ...     &    ...       \\     
 80   \ \dotfill \  &  $-$367.68 &    375.19 &  $-$3.60 &   0.99 &    4.34 &   1.09
 &   15.21 &   0.84
  &       26.70 &    ...     &        7.01 &    ...     &    ...       \\     
 81   \ \dotfill \  &  $-$388.12 &    384.99 &  $-$1.56 &   0.87 &    1.58 &   1.15
 &   15.27 &   0.84
  &       37.70 &        7.13 &        5.21 &    ...     &    ...       \\     
 82   \ \dotfill \  &  $-$401.82 &    405.92 &  $-$1.37 &   1.39 &    2.46 &   1.24
 &   15.66 &   0.94
  &       33.46 &    ...     &        5.25 &    ...     &    ...       \\     
 83   \ \dotfill \  &  $-$423.16 &    325.87 &  $-$1.34 &   1.12 &    0.49 &   1.21
 &   16.14 &   1.78
  &       88.10 &       15.43 &    ...     &    ...     &    ...       \\     
 84   \ \dotfill \  &   $-$17.07 &    136.14 &  $-$1.63 &   1.16 &    2.19 &   1.01
 &   16.43 &   1.89
  &      141.38 &        5.29 &    ...     &    ...     &    ...       \\     
 85   \ \dotfill \  &  $-$426.36 &    346.63 &  $-$3.34 &   0.89 &    4.16 &   0.84
 &   17.57 &   1.12
  &       37.96 &    ...     &        6.77 &        2.36 &    ...       \\     
 86   \ \dotfill \  &  $-$428.42 &    364.67 &  $-$3.59 &   1.42 &    4.04 &   1.20
 &   19.87 &   1.15
  &       18.83 &    ...     &        6.00 &    ...     &    ...       \\     
 87   \ \dotfill \  &  $-$402.29 &    324.52 &    0.59 &   2.04 &    1.19 &   1.33
 &   19.93 &   1.47
  &      189.65 &        7.00 &    ...     &    ...     &    ...       \\     
 88   \ \dotfill \  &  $-$401.42 &    327.64 &  $-$3.07 &   3.12 &    4.04 &   3.44
 &   20.02 &   1.47
  &    ...     &    ...     &        8.74 &        3.22 &    ...       \\     
 89   \ \dotfill \  &  $-$431.38 &    365.57 &    1.49 &   2.10 &    2.26 &   1.64
 &   20.07 &   1.68
  &       62.64 &        4.53 &    ...     &    ...     &    ...       \\     
 90   \ \dotfill \  &  $-$389.02 &    384.48 &  $-$3.34 &   2.43 &    1.73 &   1.56
 &   20.72 &   1.57
  &       44.34 &        5.52 &    ...     &    ...     &    ...       \\     
 91   \ \dotfill \  &  $-$394.69 &    385.29 &    3.36 &   2.71 &  $-$0.71 &   3.22
 &   21.83 &   1.15
  &    ...     &    ...     &       29.47 &        7.46 &    ...       \\     
 92   \ \dotfill \  &  $-$427.22 &    365.44 &  $-$4.36 &   3.05 &    3.12 &   2.33
 &   22.12 &   1.05
  &       39.54 &        6.42 &    ...     &    ...     &    ...       \\     
 93   \ \dotfill \  &  $-$413.70 &    279.51 &  $-$5.31 &   3.34 &    9.68 &   3.34
 &   22.13 &   0.94
  &    ...     &        7.02 &        5.95 &    ...     &    ...       \\     
 94   \ \dotfill \  &  $-$393.76 &    384.19 &    1.89 &   2.75 &  $-$0.94 &   3.55
 &   22.99 &   1.15
  &       75.53 &       11.65 &    ...     &    ...     &    ...       \\     
 95   \ \dotfill \  &  $-$402.97 &    324.56 &  $-$1.13 &   1.66 &    0.83 &   2.78
 &   23.91 &   2.74
  &    ...     &    ...     &       61.21 &       11.47 &    ...       \\     
 96   \ \dotfill \  &  $-$435.49 &    358.26 &  $-$3.51 &   0.93 &    3.09 &   0.99
 &   24.19 &   1.37
  &       42.87 &    ...     &        8.19 &    ...     &    ...       \\     
 97   \ \dotfill \  &  $-$402.62 &    323.02 &  $-$0.31 &   3.23 &    0.99 &   3.47
 &   24.48 &   1.57
  &       22.56 &        6.97 &    ...     &    ...     &    ...       \\     
 98   \ \dotfill \  &  $-$422.76 &    231.79 &  $-$2.72 &   1.33 &    1.35 &   1.52
 &   24.85 &   1.90
  &       49.16 &       56.67 &    ...     &    ...     &    ...       \\     
 99   \ \dotfill \  &  $-$417.88 &    358.94 &  $-$3.62 &   0.83 &    3.89 &   0.84
 &   25.29 &   1.55
  &       60.51 &       10.95 &       10.34 &    ...     &    ...       \\     
100   \ \dotfill \  &  $-$414.21 &    280.46 &  $-$3.39 &   1.00 &    5.44 &   1.33
 &   26.73 &   0.98
  &       26.63 &        4.64 &        1.51 &    ...     &    ...       \\     
101   \ \dotfill \  &  $-$422.90 &    231.69 &  $-$3.20 &   0.51 &    1.32 &   0.78
 &   27.35 &   1.68
  &       82.30 &        3.96 &       15.73 &    ...     &    ...       \\     
102   \ \dotfill \  &  $-$414.44 &    281.71 &  $-$2.51 &   0.81 &    3.70 &   1.16
 &   27.42 &   0.67
  &       15.40 &    ...     &        9.93 &        2.63 &    ...       \\     
103   \ \dotfill \  &  $-$424.43 &    226.00 &  $-$4.09 &   3.29 &    0.85 &   4.27
 &   29.03 &   0.84
  &    ...     &    ...     &        3.33 &        2.30 &    ...       \\     
104   \ \dotfill \  &  $-$422.89 &    231.03 &  $-$4.37 &   1.07 &    3.81 &   1.61
 &   30.16 &   1.15
  &       22.10 &    ...     &        2.38 &    ...     &    ...       \\     
105   \ \dotfill \  &  $-$424.90 &    203.86 &  $-$4.95 &   1.07 &    1.76 &   1.10
 &   32.26 &   0.63
  &       18.90 &    ...     &        1.98 &    ...     &    ...       \\     
106   \ \dotfill \  &   $-$76.82 &   $-$40.30 &  $-$2.14 &   0.80 &    0.11 &   0.78
 &   33.95 &   1.40
  &       25.26 &        3.63 &        6.78 &    ...     &    ...       \\     
107   \ \dotfill \  &  $-$419.15 &    224.97 &  $-$4.49 &   2.59 &    3.21 &   2.99
 &   35.42 &   2.32
  &    ...     &        3.32 &        9.05 &    ...     &    ...       \\     
108   \ \dotfill \  &  $-$425.23 &    263.61 &  $-$3.21 &   0.19 &    3.97 &   0.29
 &   36.75 &   3.21
  &    ...     &        7.23 &       32.30 &        4.01 &      116.30   \\     
109   \ \dotfill \  &  $-$428.83 &    214.15 &  $-$3.85 &   0.76 &    9.80 &   1.29
 &   41.94 &   1.90
  &       28.00 &    ...     &       17.86 &        1.86 &    ...       \\     
110   \ \dotfill \  &  $-$428.82 &    215.68 &  $-$3.92 &   0.26 &    1.16 &   0.61
 &   44.43 &   2.32
  &      256.34 &       62.45 &      115.72 &    ...     &    ...       \\     
111   \ \dotfill \  &  $-$370.82 &    229.75 &  $-$4.01 &   2.72 &    2.79 &   3.76
 &   49.11 &   1.15
  &    ...     &    ...     &        3.01 &        2.72 &    ...       \\     
112   \ \dotfill \  &  $-$409.02 &    230.41 &  $-$3.60 &   0.96 &    2.22 &   1.03
 &   52.89 &   1.58
  &    ...     &    ...     &       12.79 &        8.28 &    ...       \\     
113   \ \dotfill \  &  $-$423.54 &    214.87 &  $-$1.41 &   3.42 &    0.92 &   3.09
 &   54.79 &   0.37
  &       10.90 &        1.56 &    ...     &    ...     &    ...       \\     
114   \ \dotfill \  &  $-$373.63 &    228.51 &    0.40 &   3.59 &    8.19 &   2.85
 &   55.71 &   1.89
  &    ...     &        2.69 &        8.62 &    ...     &    ...       \\     
115   \ \dotfill \  &  $-$421.19 &    222.93 &  $-$3.91 &   0.58 &    3.24 &   0.65
 &   55.79 &   1.47
  &       26.49 &    ...     &    ...     &        7.40 &       36.98   \\     
116   \ \dotfill \  &  $-$386.01 &    236.21 &  $-$4.54 &   0.47 &    4.13 &   0.66
 &   57.96 &   1.12
  &    ...     &    ...     &        2.11 &        2.49 &       59.33   \\     
117   \ \dotfill \  &  $-$420.92 &    222.06 &  $-$5.31 &   1.66 &    3.20 &   1.56
 &   58.81 &   1.75
  &       15.24 &        3.14 &       13.45 &    ...     &    ...       \\     
118   \ \dotfill \  &  $-$383.90 &    232.24 &  $-$3.19 &   3.54 &    3.72 &   4.52
 &   59.37 &   1.17
  &    ...     &    ...     &        5.76 &        2.13 &    ...       \\     
119   \ \dotfill \  &  $-$375.03 &    227.21 &  $-$5.16 &   2.32 &    2.25 &   2.07
 &   61.86 &   1.06
  &    ...     &        2.88 &       15.01 &        1.78 &    ...       \\     
120   \ \dotfill \  &  $-$425.92 &    215.03 &  $-$2.17 &   0.40 &    0.67 &   0.48
 &   62.05 &   1.05
  &    ...     &        2.62 &       16.69 &    ...     &       18.60   \\     
121   \ \dotfill \  &  $-$420.55 &    223.46 &  $-$3.02 &   0.30 &    1.51 &   0.62
 &   64.05 &   3.72
  &     1701.62 &       27.69 &       21.30 &    ...     &    ...       \\     
122   \ \dotfill \  &  $-$422.97 &    216.81 &  $-$4.24 &   0.87 &    2.96 &   1.29
 &   67.83 &   2.11
  &       18.70 &        7.35 &       11.55 &    ...     &    ...       \\     
123   \ \dotfill \  &  $-$421.18 &    214.93 &  $-$4.60 &   0.83 &    0.86 &   1.05
 &   69.41 &   2.95
  &      124.73 &        9.66 &        4.63 &    ...     &    ...       \\     
\hline
\end{tabular}
}

\footnotemark[1] Water maser features detected 
toward W51 North. The feature is designated as W~51N:I2002 {\it N}, 
where {\it N} is the ordinal source number given in this column (I2002 
stands for sources found by Imai et al. and listed in 2002).  \\
\footnotemark[2] Relative value with respect to the location of the position-reference 
maser feature: W51N:I2002 {\it 4}. \\
\footnotemark[3] Relative value with respect to the motion of the position-reference 
maser feature: W51N:I2002 {\it 4}. \\
\footnotemark[4] Relative value with respect to the assumed systemic velocity of 
$V_{\mbox{LSR}}=$ 56.0 km s$^{-1}$. \\
\footnotemark[5] Peak intensity at five epochs. \\
 
\end{table*}

\begin{table*}[ht]
\caption{Same as table 2 but toward W51 Main.}\label{tab:pmotionsM}
{\scriptsize
\begin{tabular}{l@{ }r@{ \ }rr@{ \ }rr@{ \ }rrrr@{ \ }r@{ \ }r@{ \ }r@{ \ }r} \hline \hline
& \multicolumn{2}{c}{Offset\footnotemark[2]}
 & \multicolumn{4}{c}{Proper motion\footnotemark[3]}
 & \multicolumn{2}{c}{Radial motion\footnotemark[4]}
 & \multicolumn{5}{c}{Peak intensity\footnotemark[5]} \\
Maser Feature\footnotemark[1] & \multicolumn{2}{c}{(mas)} 
 & \multicolumn{4}{c}{(mas yr$^{-1}$)}
 & \multicolumn{2}{c}{(km s$^{-1}$)}
 & \multicolumn{5}{c}{(Jy beam$^{-1}$)} \\
 & \multicolumn{2}{c}{\ \hrulefill \ } 
 & \multicolumn{4}{c}{\ \hrulefill \ } 
 & \multicolumn{2}{c}{\ \hrulefill \ } 
 & \multicolumn{5}{c}{\ \hrulefill \ } \\                                        
(W51M:I2002) & R.A. & decl. 
 & $\mu_{x}$ & $\sigma \mu_{x}$ & $\mu_{y}$ 
 & $\sigma \mu_{y}$ & $V_{z}$ & $\Delta V_{z}$
 & j99052 & j990092 & j99126 & j99156 & j99294 \\ \hline
  1   \ \dotfill \  &   1289.70 & $-$1350.62 &    9.13 &   1.18 &  $-$0.51 &   2.29
 & $-$92.63 &   0.94 
  &       38.07 &    ...     &        2.76 &    ...     &    ...       \\     
  2   \ \dotfill \  &  $-$131.31 &    149.12 &    0.65 &   1.39 &    0.93 &   1.17
 & $-$22.52 &   0.63 
  &       19.50 &    ...     &        4.85 &    ...     &    ...       \\     
  3   \ \dotfill \  &   $-$42.42 &   $-$77.73 &    2.71 &   9.79 &    2.07 &   8.70
 &    1.13 &   1.58 
  &    ...     &    ...     &      696.00 &      124.63 &    ...       \\     
  4   \ \dotfill \  &   $-$28.80 &   $-$41.18 &    1.77 &   1.43 &    1.11 &   1.05
 &    2.33 &   1.37 
  &     3592.00 &     1213.00 &    ...     &    ...     &    ...       \\     
  5   \ \dotfill \  &   $-$28.79 &   $-$41.63 &    0.40 &   0.56 &    2.56 &   0.53
 &    2.41 &   1.15 
  &    ...     &    ...     &    ...     &      174.50 &     2312.00   \\     
  6   \ \dotfill \  &      4.98 &     82.41 &    0.69 &   0.89 &    3.48 &   1.19
 &    4.17 &   1.05 
  &      182.00 &       46.69 &       39.00 &    ...     &    ...       \\     
  7   \ \dotfill \  &    $-$6.59 &   $-$10.75 &    1.76 &   0.23 &    2.60 &   0.65
 &    4.26 &   1.05 
  &    ...     &       83.63 &       80.00 &       12.28 &      271.00   \\     
  8   \ \dotfill \  &    $-$7.00 &    $-$9.13 &    1.62 &   0.29 &    2.84 &   0.38
 &    4.38 &   1.12 
  &    ...     &    ...     &      115.00 &       32.75 &      176.00   \\     
  9   \ \dotfill \  &    $-$2.57 &   $-$11.88 &    1.34 &   2.58 &    5.98 &   2.00
 &    5.29 &   1.37 
  &    ...     &       33.59 &       62.38 &    ...     &    ...       \\     
 10   \ \dotfill \  &     23.96 &    130.94 &    2.00 &   1.06 &    3.43 &   1.04
 &    8.29 &   2.11 
  &      132.50 &    ...     &      161.25 &      101.04 &    ...       \\     
 11   \ \dotfill \  &    $-$1.02 &   $-$16.00 &    5.77 &   4.78 &  $-$0.79 &   6.76
 &    9.82 &   0.94 
  &    ...     &        8.94 &       12.72 &    ...     &    ...       \\     
 12   \ \dotfill \  &     24.13 &    131.12 &    2.50 &   0.57 &    3.87 &   0.91
 &   10.91 &   2.11 
  &       74.66 &    ...     &       51.69 &    ...     &    ...       \\     
 13   \ \dotfill \  &     23.64 &    130.53 &  $-$2.18 &  11.63 &  $-$1.13 &   9.89
 &   12.27 &   1.26 
  &    ...     &    ...     &       47.13 &        6.45 &    ...       \\     
 14   \ \dotfill \  &    $-$1.10 &    $-$7.21 &    1.19 &   2.75 &    2.59 &   3.49
 &   12.44 &   0.84 
  &    ...     &       19.28 &       40.88 &    ...     &    ...       \\     
 15   \ \dotfill \  &      0.00 &      0.00 &    2.30 &   0.26 &    2.99 &   0.26
 &   14.65 &   2.49 
  &      562.00 &      130.75 &      456.00 &       73.30 &      409.00   \\     
 16   \ \dotfill \  &      0.70 &    $-$0.11 &    2.23 &   3.79 &    4.50 &   2.57
 &   17.20 &   1.68 
  &    ...     &    ...     &      110.75 &       16.17 &    ...       \\     
 17   \ \dotfill \  &      0.87 &    $-$0.10 &    2.41 &   1.77 &    1.33 &   1.35
 &   17.45 &   1.37 
  &      161.00 &       38.75 &    ...     &    ...     &    ...       \\     
 18   \ \dotfill \  &    $-$0.82 &    $-$0.53 &    2.22 &   0.22 &    3.36 &   0.24
 &   18.59 &  42.81 
  &      140.37 &    ...     &      139.75 &       47.13 &      654.25   \\     
 19   \ \dotfill \  &   $-$32.66 &  $-$155.42 &   10.32 &   4.32 &  $-$0.26 &   3.37
 &   18.77 &   1.37 
  &       47.69 &       42.75 &    ...     &    ...     &    ...       \\     
 20   \ \dotfill \  &     36.65 &   $-$73.17 &    2.25 &   0.21 &    4.43 &   0.21
 &   21.05 &   1.58 
  &      171.00 &       33.34 &    ...     &       22.16 &      159.50   \\     
 21   \ \dotfill \  &   $-$34.35 &  $-$168.20 &    1.14 &   1.43 &    1.95 &   1.42
 &   23.18 &   1.61 
  &    ...     &       14.70 &      111.75 &        7.66 &    ...       \\     
 22   \ \dotfill \  &     32.53 &   $-$99.70 &    2.15 &   1.56 &    1.52 &   1.43
 &   25.63 &   1.15 
  &      214.50 &       21.06 &    ...     &    ...     &    ...       \\     
 23   \ \dotfill \  &   $-$11.94 &  $-$143.57 &  $-$0.23 &   2.64 &    2.66 &   2.22
 &   25.94 &   1.15 
  &       99.75 &        9.56 &    ...     &    ...     &    ...       \\     
 24   \ \dotfill \  &     38.07 &   $-$98.60 &    7.76 &   2.17 &    3.70 &   2.32
 &   27.24 &   1.05 
  &       71.00 &       12.44 &    ...     &    ...     &    ...       \\     
 25   \ \dotfill \  &   $-$11.78 &  $-$144.06 &  $-$3.52 &   6.52 &    6.51 &   6.17
 &   27.67 &   1.26 
  &    ...     &    ...     &       13.63 &       14.69 &    ...       \\     
 26   \ \dotfill \  &    118.35 &  $-$168.44 &    0.97 &   0.47 &    4.87 &   0.56
 &   27.78 &   2.22 
  &    ...     &    ...     &       40.06 &    ...     &       76.31   \\     
 27   \ \dotfill \  &    $-$8.64 &  $-$141.69 &    1.20 &   0.92 &    1.88 &   0.88
 &   27.95 &   1.19 
  &      236.50 &       16.22 &       23.97 &    ...     &    ...       \\     
 28   \ \dotfill \  &   $-$18.30 &  $-$151.66 &    3.97 &   2.44 &    1.72 &   2.40
 &   30.31 &   1.05 
  &       18.00 &        7.23 &    ...     &    ...     &    ...       \\     
 29   \ \dotfill \  &     30.92 &  $-$104.23 &    1.77 &   0.36 &    4.34 &   0.47
 &   32.82 &   1.16 
  &       45.13 &       16.31 &    ...     &        9.12 &      202.00   \\     
 30   \ \dotfill \  &   $-$15.07 &   $-$66.38 &    2.80 &   0.35 &    3.12 &   0.48
 &   32.89 &   1.16 
  &    ...     &    ...     &       16.41 &    ...     &      132.75   \\     
 31   \ \dotfill \  &   $-$24.22 &   $-$66.13 &    1.54 &   0.68 &    2.44 &   0.94
 &   33.65 &   1.26 
  &       35.75 &    ...     &       32.44 &    ...     &    ...       \\     
 32   \ \dotfill \  &     32.54 &  $-$106.43 &    0.75 &   1.41 &    4.15 &   1.20
 &   36.84 &   0.91 
  &      100.00 &       14.81 &    ...     &        4.47 &    ...       \\     
 33   \ \dotfill \  &     34.57 &  $-$106.32 &    1.64 &   0.35 &    5.79 &   0.32
 &   38.88 &   1.79 
  &      325.00 &       92.50 &    ...     &       16.63 &      303.00   \\     
 34   \ \dotfill \  &   $-$30.99 &  $-$167.34 &  $-$1.27 &   2.51 &    3.28 &   1.70
 &   39.76 &   1.54 
  &    ...     &       36.81 &       28.94 &        7.16 &    ...       \\     
 35   \ \dotfill \  &     38.09 &  $-$107.99 &    4.29 &   2.37 &    3.95 &   1.52
 &   41.31 &   1.05 
  &       86.62 &       40.37 &    ...     &    ...     &    ...       \\     
 36   \ \dotfill \  &     37.25 &  $-$108.55 &    2.47 &   3.75 &    4.03 &   9.63
 &   42.74 &   2.42 
  &      107.00 &      107.50 &    ...     &    ...     &    ...       \\     
 37   \ \dotfill \  &     37.11 &  $-$109.39 &    0.76 &   1.34 &    4.42 &   1.15
 &   43.63 &   2.42 
  &    ...     &    ...     &    ...     &        8.94 &     1278.00   \\     
 38   \ \dotfill \  &     40.77 &     87.07 &    0.13 &   1.92 &    4.60 &   1.61
 &   46.48 &   1.37 
  &    ...     &       22.81 &       37.12 &    ...     &    ...       \\     
 39   \ \dotfill \  &     28.23 &  $-$115.36 &    2.37 &   2.83 &    2.89 &   2.26
 &   50.73 &   0.84 
  &       25.25 &        4.34 &    ...     &    ...     &    ...       \\     
 40   \ \dotfill \  &   $-$26.72 &  $-$164.19 &    6.17 &   4.69 &    3.11 &   3.71
 &   52.64 &   1.47 
  &       16.80 &        7.25 &    ...     &    ...     &    ...       \\     
 41   \ \dotfill \  &   $-$27.11 &  $-$162.25 &    0.48 &   1.09 &    4.20 &   1.68
 &   56.36 &   1.97 
  &       69.50 &       19.69 &        7.72 &    ...     &    ...       \\     
 42   \ \dotfill \  &     39.75 &     40.68 &    0.47 &   1.00 &    5.50 &   1.34
 &   59.05 &   1.62 
  &       22.37 &       12.44 &       12.02 &    ...     &    ...       \\     
 43   \ \dotfill \  &   $-$28.60 &   $-$81.84 &    7.03 &   1.07 &    6.66 &   0.68
 &   61.16 &   1.61 
  &      106.00 &       22.88 &       12.37 &    ...     &    ...       \\     
 44   \ \dotfill \  &    $-$2.27 &  $-$140.11 &    4.19 &   3.73 &    1.43 &   2.32
 &   63.03 &   1.37 
  &      119.00 &        4.66 &    ...     &    ...     &    ...       \\     
 45   \ \dotfill \  &     32.90 &   $-$91.38 &    0.87 &   0.57 &    3.63 &   0.71
 &   65.62 &   1.15 
  &    ...     &    ...     &        4.87 &    ...     &      209.00   \\     
 46   \ \dotfill \  &     14.72 &  $-$142.01 &    1.17 &   0.96 &    1.54 &   1.09
 &   67.79 &   2.00 
  &      111.00 &       35.75 &       33.94 &        4.16 &    ...       \\     
 47   \ \dotfill \  &     27.03 &  $-$370.19 &    6.89 &   3.47 &    1.30 &   2.21
 &   80.99 &   1.48 
  &    ...     &       11.19 &       52.75 &    ...     &    ...       \\     
 48   \ \dotfill \  &     29.61 &  $-$377.16 &  $-$4.22 &   3.30 &    7.13 &   3.76
 &  101.63 &   0.84 
  &    ...     &        8.53 &        8.56 &    ...     &    ...       \\   
\hline
\end{tabular}
}

\footnotemark[1] Water maser features detected toward W~51 Main. 
The feature is designated as W51M:I2002 {\it N}, where {\it N} is the 
ordinal source number given in this column (I2002 stands for sources 
found by Imai et al. and listed in 2002).  \\
\footnotemark[2] Relative value with respect to the location of the position-reference 
maser feature: W51M:I2002 {\it 15}. \\
\footnotemark[3] Relative value with respect to the motion of the position-reference 
maser feature: W51N:I2002 {\it 4}. \\
\footnotemark[4] Relative value with respect to the assumed systemic velocity of 
$V_{\mbox{LSR}}=$ 55.0 km s$^{-1}$. \\
\footnotemark[5] Peak intensity at five epochs. \\
\end{table*}

\begin{table*}[ht]
\caption{Same as table 2 but toward W51 South (e4).}\label{tab:pmotionsS}
{\scriptsize
\begin{tabular}{l@{ }r@{ \ }rr@{ \ }rr@{ \ }rrrr@{ \ }r@{ \ }r@{ \ }r@{ \ }r} \hline \hline
& \multicolumn{2}{c}{Offset\footnotemark[2]}
 & \multicolumn{4}{c}{Proper motion\footnotemark[3]}
 & \multicolumn{2}{c}{Radial motion\footnotemark[4]}
 & \multicolumn{5}{c}{Peak intensity\footnotemark[5]} \\
Maser Feature\footnotemark[1] & \multicolumn{2}{c}{(mas)} 
 & \multicolumn{4}{c}{(mas yr$^{-1}$)}
 & \multicolumn{2}{c}{(km s$^{-1}$)}
 & \multicolumn{5}{c}{(Jy beam$^{-1}$)} \\
 & \multicolumn{2}{c}{\ \hrulefill \ } 
 & \multicolumn{4}{c}{\ \hrulefill \ } 
 & \multicolumn{2}{c}{\ \hrulefill \ } 
 & \multicolumn{5}{c}{\ \hrulefill \ } \\                                        
(W51S:I2002) & R.A. & decl. 
 & $\mu_{x}$ & $\sigma \mu_{x}$ & $\mu_{y}$ 
 & $\sigma \mu_{y}$ & $V_{z}$ & $\Delta V_{z}$
 & j99052 & j990092 & j99126 & j99156 & j99294 \\ \hline
  1   \ \dotfill \  &    281.54 &  $-$261.33 &    2.03 &   4.18 &    1.84 &   7.78
 &$-$106.14 &   1.05 
  &    ...     &    ...     &       15.10 &        3.41 &    ...       \\     
  2   \ \dotfill \  &    259.42 &  $-$255.53 &    2.56 &   1.46 &    0.55 &   2.15
 & $-$91.88 &   1.89 
  &    ...     &    ...     &       39.82 &       12.82 &    ...       \\     
  3   \ \dotfill \  & $-$1049.26 &  $-$145.00 &  $-$5.00 &   1.85 &    0.35 &   2.41
 &  $-$7.92 &   1.68 
  &    ...     &    ...     &      383.00 &        6.41 &    ...       \\     
  4   \ \dotfill \  &   $-$29.80 &    $-$6.27 &  $-$1.82 &   0.28 &    2.43 &   0.29
 &  $-$7.26 &   1.68 
  &    ...     &    ...     &      352.75 &       51.22 &      825.00   \\     
  5   \ \dotfill \  &    $-$2.93 &      1.73 &  $-$2.28 &   1.18 &    2.58 &   1.00
 &    0.80 &   1.19 
  &      210.00 &    ...     &       60.53 &        6.88 &    ...       \\     
  6   \ \dotfill \  &      0.00 &      0.00 &  $-$1.63 &   0.55 &    2.38 &   0.33
 &    2.01 &   1.51 
  &      275.00 &      139.69 &      108.37 &       16.27 &      390.50   \\     
  7   \ \dotfill \  &  $-$323.67 &     37.11 &  $-$1.85 &   0.55 &    1.84 &   0.52
 &    2.50 &   0.70 
  &       99.50 &    ...     &       28.38 &        5.67 &    ...       \\     
  8   \ \dotfill \  &  $-$315.17 &     36.83 &  $-$0.57 &   0.79 &    2.74 &   0.77
 &    3.99 &   0.74 
  &       38.78 &    ...     &    ...     &        5.01 &    ...       \\     
  9   \ \dotfill \  &  $-$269.53 &     17.89 &  $-$2.16 &   0.20 &    2.29 &   0.19
 &    4.81 &   1.55 
  &     1016.00 &    ...     &      494.00 &       70.31 &    ...       \\     
 10   \ \dotfill \  &  $-$828.27 &    572.16 &    2.41 &   5.71 &    3.26 &   3.58
 &   15.00 &   1.37 
  &    ...     &    ...     &       38.56 &        6.78 &    ...       \\  
\hline
\end{tabular}
}

\footnotemark[1] Water maser features detected toward W~51 South. 
The feature is designated as W51S:I2002 {\it N}, where {\it N} is the ordinal 
source number given in this column (I2002 stands for sources found by 
Imai et al. and listed in 2002). \\
\footnotemark[2] Relative value with respect to the location of the position-reference 
maser feature: W51S:I2002 {\it 6}. \\
\footnotemark[3] Relative value with respect to the motion of the position-reference 
maser feature: W51N:I2002 {\it 4}. \\
\footnotemark[4] Relative value with respect to the assumed systemic velocity of 
$V_{\mbox{LSR}}=$ 59.0 km s$^{-1}$. \\
\footnotemark[5] Peak intensity at five epochs. \\
\end{table*}

\begin{table*}[p]
\caption{Best-fit models for the maser velocity field of 
the W51 North outflow}
\label{tab:w51model-fit}

\begin{center}
\begin{tabular}{lr@{}c@{$\pm$}lr@{}c@{$\pm$}l} 
\hline \hline
\multicolumn{1}{c}{Parameter}& \multicolumn{3}{c}{Step 1\footnotemark[1] } 
& \multicolumn{3}{c}{Step 2\footnotemark[2] } \\
\hline \multicolumn{7}{c}{Offsets} \\ \hline
\multicolumn{7}{l}{Velocity:} \\
\hspace*{10pt} $V_{\mbox{0x}}$\footnotemark[3] (\kms) \dotfill \ 
&$-$52 & & 18 & $-$50 & & 19 \\  
\hspace*{10pt} $V_{\mbox{0y}}$\footnotemark[3] (\kms) \dotfill \ 
& 22 & & 12 & 14 & & 34 \\
\hspace*{10pt} $V_{\mbox{0z}}$\footnotemark[4] (\kms) \dotfill \ 
& \multicolumn{3}{c}{0\footnotemark[5]} & 2 & & 7 \\
\multicolumn{7}{l}{Position:} \\
\hspace*{10pt} $x_{\mbox{0}}$ (arcsec) \dotfill \ & $-$0.28 & & 0.06 
& $-$0.23 & & 0.08 \\
\hspace*{10pt} $y_{\mbox{0}}$ (arcsec) \dotfill \  & 0.11 & & 0.05 
& 0.13 & & 0.10 \\
\hline \multicolumn{7}{c}{Velocity field} \\ \hline
\multicolumn{7}{l}{Radial outflow:} \\
\hspace*{10pt} $V_{\mbox{1}}$ (\kms arcsec$^{-\alpha}$) \dotfill \ 
& \multicolumn{3}{c}{...\footnotemark[6]}
& 27 & & 6 \\
\hspace*{10pt} $\alpha$ \dotfill \ 
& \multicolumn{3}{c}{...\footnotemark[6]} & $-$0.9 & & 0.2 \\
Distance $d$ (kpc) \dotfill \ & \multicolumn{3}{c}
{6.0\footnotemark[7]} & 6.1 & & 1.3 \\ \hline
RMS residual $\sqrt{S^{2}}$ \dotfill \ &
\multicolumn{3}{c}{16.23} & \multicolumn{3}{c}{3.67} \\ \hline 
\multicolumn{7}{c}{} \\
\end{tabular}
\end{center}

\footnotemark[1] Assuming independent expansion velocity of maser 
features. \\
\footnotemark[2] Asumming a common expansion velocity as a function of 
distance from the outflow origin as shown in the main text. \\
\footnotemark[3] Relative value with respect to the position-reference 
maser feature. \\ 
\footnotemark[4] Relative value with respect to \vlsr$\equiv $ 56\kms. \\ 
\footnotemark[5] Step 1 assumes the systemic radial velocity:  
$V_{\mbox{0z}}\equiv $ 0.0\kms. \\ 
\footnotemark[6] The solution determines a radial outflow velocity 
$V_{\mbox{exp}}$(i) independently for each feature with a proper motion. \\
\footnotemark[7] Distance is completely covariant with the $z_{\mbox{i}}$ 
and $V_{\mbox{exp}}$(i) and cannot be determined: $d\equiv$7.0 kpc.
\end{table*}

\begin{table*}[p]
\caption{Best-fit model for the maser velocity field of 
the W51 South (e4) outflow}
\label{tab:w51smodel-fit}

\begin{center}
\begin{tabular}{lr@{}c@{$\pm$}lr} \hline \hline
\multicolumn{4}{l}{Velocity offset:} \\ \hline
\hspace*{10pt} $V_{\mbox{0x}}$\footnotemark[1] (\kms) \dotfill \ & 83 & & 15 \\  
\hspace*{10pt} $V_{\mbox{0y}}$\footnotemark[1] (\kms) \dotfill \ & 71 & & 8 \\
\hspace*{10pt} $V_{\mbox{0z}}$\footnotemark[2] (\kms) \dotfill \ 
& \multicolumn{3}{c}{0\footnotemark[3]} \\ \hline
\multicolumn{4}{l}{Position offset:} \\ \hline
\hspace*{10pt} $x_{\mbox{0}}$ (arcsec) \dotfill \ & 0.23 & & 0.14 \\
\hspace*{10pt} $y_{\mbox{0}}$ (arcsec) \dotfill \  & $-$0.01 & & 0.02 \\
Distance $d$ (kpc) \dotfill \ & \multicolumn{3}{c}{6.0\footnotemark[4]} \\ \hline
RMS residual $\sqrt{S^{2}}$ \dotfill \ & \multicolumn{3}{c}{9.32} \\ \hline 
\multicolumn{4}{c}{} \\
\end{tabular}
\end{center}

\footnotemark[1] Relative value with respect to the position-reference 
maser feature in W~51N (W51N: I2002 {\it 4}). \\ 
\footnotemark[2] Relative value with respect to \vlsr$\equiv $ 
59\kms. \\ 
\footnotemark[3] Assuming the systemic radial velocity:  
$V_{\mbox{0z}}\equiv $ 0.0\kms. \\ 
\footnotemark[4] Distance is completely covariant with the $z_{\mbox{i}}$ 
and $V_{\mbox{exp}}$(i) and cannot be determined: $d\equiv$6.0 kpc.
\end{table*}

\begin{table*}[ht]
\caption{Eigenvalues and eigenvectors of velocity variance--covariance 
matrixes of the maser kinematics in W51 North and Main after 
diagonalization.}\label{tab:vvcm}
\begin{center}
\begin{tabular}{llrrr}\hline \hline
& Eigenvalue & & & \\
& \hspace*{\fill}(km$^{2}$s$^{-2}$) 
& \multicolumn{3}{c}{Normalized eigenvector} \\ \hline
W51N\dotfill & 4759 & ($-$0.6332, & 0.7141, & 0.2984) \\
& 2084 & (0.6819, & 0.6971, & $-$0.2213) \\
& 406 & (0.3660, & $-$0.0634, & 0.9284) \\ \hline
W51M\dotfill & 7169 & (0.9214, & $-$0.3703, & $-$0.1182) \\
& 2381 & (0.3887, & 0.8778, & 0.2800) \\
& 653 & (0.0001, & $-$0.3039, & 0.9527) \\ \hline
\end{tabular}
\end{center}
\end{table*}

\begin{table*}[ht]
\caption{Relative bulk motions of three maser clusters with respect 
to the position-reference feature in W\ 51 North.}
\label{tab:rmotion}
\begin{center}
\begin{tabular}{p{150pt}cccl} \hline \hline
& $V_{\mbox{X}}$ & $V_{\mbox{Y}}$ 
& $V_{\mbox{Z}}$\footnotemark[1] & Used \\
& (\kms) & (\kms) & (\kms) & proper motions \\ \hline
W51 North (average) \dotfill \
& $-$52$\pm$54\footnotemark[2] & 36$\pm$58\footnotemark[2]  
& 60$\pm$35\footnotemark[2]  & 92 \\
\hspace*{40pt} (model fit) \dotfill \ 
& $-$50$\pm$19 & 14$\pm$34 & 58$\pm$7 & 68 \\
W51 Main (e2) (average) \dotfill \
& 61$\pm$63\footnotemark[2]  & 91$\pm$45\footnotemark[2]  
& 70$\pm$40\footnotemark[2]  & 37 \\
W51 South (e4)  (average) \dotfill \
& $-$12$\pm$66\footnotemark[2] & 61$\pm$25\footnotemark[2] 
& 29$\pm$54\footnotemark[2] & 10 \\ 
\hspace*{40pt} (model fit) \dotfill \
& 71$\pm$15 & 59$\pm$8 & 59\footnotemark[3] & 10 \\ \hline
\multicolumn{5}{c}{ } \\
\end{tabular}
\end{center}

\footnotemark[1] Velocity with respect to the local standard of rest. \\
\footnotemark[2] Uncertainty was estimated from the standard deviation 
of motions. \\
\footnotemark[3] Assuming \vlsr$\equiv$ 59\kms. 
\end{table*}

\end{document}